\documentclass[numberedappendix, appendixfloats, twocolappendix, nofootinbib,onecolbib,square, comma, sort&compress, numbers]{emulateapj}
\usepackage{amsmath}
\usepackage{comment}
\usepackage{threeparttable}
\usepackage{subfigure}
\usepackage{graphicx}
\usepackage{hyperref}
\usepackage{CJKutf8}
\usepackage{pbox}
\usepackage[all]{hypcap} 
\usepackage{tikz}
\usetikzlibrary{shapes.geometric, arrows}
\tikzstyle{decision} = [diamond, minimum width=1cm, minimum height=0cm, text centered, draw=black,aspect=2]
\tikzstyle{process} = [rectangle, minimum width=1cm, minimum height=0.5cm, text centered, draw=black]
\tikzstyle{line} = [draw, -latex']
\tikzstyle{io} = [trapezium, trapezium left angle=70, trapezium right angle=110, minimum width=1cm, minimum height=1cm, text centered, draw=black]
\tikzstyle{continue} = [rectangle, minimum width=1cm, minimum height=1cm]

\begin{document}


\newcommand{\kcbias}{$(4.69\pm{3.46})\%$ } 
\newcommand{\alphakc}{$(3.98\pm{3.22})\times 10^{-6}\mathrm{K^{-1}}$ }  

\newcommand{\kcbiasnilc}{$(6.76\pm{3.74})\%$ } 
\newcommand{\alphakcnilc}{$(7.28\pm{1.55})\times 10^{-6}\mathrm{K^{-1}}$ }

\newcommand{\pcbias}{$(15.60\pm{3.08})\%$ }
\newcommand{\alphapc}{$(5.66\pm{1.23})\times 10^{-6}\mathrm{K^{-1}}$ }

\newcommand{\pcbiasnilc}{$(16.69\pm{3.52})\%$ }
\newcommand{\alphapcnilc}{$(5.64\pm{0.81})\times 10^{-6}\mathrm{K^{-1}}$ }

\newcommand{\kcbiasn}{$4.22\pm{3.18}$ }  
\newcommand{\alphakcn}{$3.95\pm{3.24}$ }

\newcommand{\kcbiasnilcn}{$6.75\pm{3.53}$ }
\newcommand{\alphakcnilcn}{$7.28\pm{1.55}$ }

\newcommand{\pcbiasn}{$15.60\pm{3.08}$ }
\newcommand{\alphapcn}{$5.66\pm{1.23}$ }

\newcommand{\pcbiasnilcn}{$16.69\pm{3.52}$ }
\newcommand{\alphapcnilcn}{$5.64\pm{0.81}$ }

\newcommand{\pcbiasp}{$(6.93\pm{2.62})\%$ }
\newcommand{\alphapcp}{$(1.22\pm{1.90}) \times 10^{-6}K^{-1}$}


\newcommand{\kcbiasnp}{$2.28\pm{3.69}$ } 
\newcommand{\alphakcnp}{$4.23\pm{2.37}$ }  

\newcommand{\kcbiasnilcnp}{$3.92\pm{3.93}$ } 
\newcommand{\alphakcnilcnp}{$6.53\pm{2.34}$}

\newcommand{\pcbiasnp}{$15.63\pm{3.15}$ }
\newcommand{\alphapcnp}{$5.68\pm{2.62}$}

\newcommand{\pcbiasnilcnp}{$16.54\pm{3.09}$ }
\newcommand{\alphapcnilcnp}{$5.61\pm{2.57}$}

\shorttitle{Contamination in tSZ Signal}
\shortauthors{Yan et al.}

\title{An Assessment of Contamination in the thermal-SZ Map Using cross-correlations}

\author{
Ziang~Yan\begin{CJK*}{UTF8}{gbsn}(颜子昂)\altaffilmark{1},Alireza~Hojjati\altaffilmark{1},Tilman~Tröster\altaffilmark{2},Gary~Hinshaw\altaffilmark{1},Ludovic~van Waerbeke\altaffilmark{1}\end{CJK*}
}

\email{yanza15@phas.ubc.ca}

\altaffiltext{1}{Department of Physics and Astronomy, University of British Columbia, Vancouver, BC, V6T 1Z1, Canada} 
\altaffiltext{2}{Institute for Astronomy, University of Edinburgh, Royal Observatory, Blackford Hill,
Edinburgh, EH9 3HJ, UK}

\begin{abstract}
We search for potential galactic and extragalactic dust contamination in thermal Sunyaev-Zeldovich (tSZ) maps derived from the Planck data.  To test for contamination, we apply a variety of galactic dust and cosmic infrared background (CIB) models to the data as part of the $y$ map reconstruction process. We evaluate the level of contamination by cross-correlating these $y$ maps with mass tracers based on weak lensing data.  The lensing data we use are the convergence map, $\kappa$, from the Red Sequence Cluster Lensing survey (RCSLens), and the CMB lensing potential map, $\phi$, from the Planck Collaboration. We make a CIB-subtracted $y$ map and measure the cross-correlation between it and the lensing data. By comparing it with CIB-contaminated cross-correlation, we find that the cross-correlation between $\kappa$ and $y$ is only slightly contaminated by CIB signal, at the level of $6.8 \pm {3.5}$ \%, which implies that previous detections of $\kappa \, \times \, y$ are robust to CIB contamination.  However, we find that $\phi \, \times \, y$ is more significantly contaminated, by $16.7 \pm {3.5}$ \%, because the CMB lensing potential probes higher redshift sources that overlap more with the CIB sources.  We find that Galactic dust does not significantly contaminate either cross-correlation signal.

\end{abstract}


 \section{Introduction}

The thermal Sunyaev-Zeldovich effect \citet{zeldovich1969interaction} is the inverse Compton scattering of Cosmic Microwave Background (CMB) photons by high energy electrons. 
CMB photons get an energy boost through this effect and their energy spectrum is distorted. This effect provides a useful tool to observe distant structures where ionized gas is present \citep[e.g., galaxy groups, clusters, or filaments; see][]{planck2014plancksz,tanimura2017search}. It mainly occurs in the hot intracluster gas in galaxy clusters. The effect is independent of redshift because it is a scattering effect, thus high redshift clusters can be observed more easily than with redshift dependent signals such as X-rays or optical emission. Besides searching for new clusters, the tSZ effect can also be used to constrain cosmological parameters by providing information on the abundance of galaxy clusters, which depends on $\sigma_8$ and $\Omega_m$ \citet{molnar2002constraints}.\par 
 
With the current observational precision, it is possible to detect the tSZ signal from galaxy clusters \citep[e.g.,][]{hasselfield2013atacama} after filtering out other components like the CMB, galactic dust, and point sources. Moreover, since the frequency dependence of the tSZ is well understood, it is possible to extract the dimensionless Comptonization parameter, $y$, from multi-frequency sky maps. In 2015, the Planck team constructed two full-sky tSZ  maps \citet{aghanim2016planck} from Planck data using frequency channels from 30 to 857 GHz, with two distinct component separation algorithms, NILC \citep[Needlet Internal Linear Combination method,][]{delabrouille2009full} and MILCA \citep[Modified Internal Linear Combination Algorithm,][]{hurier2013milca}. Several subsequent analyses have used these tSZ  maps \citep[e.g.,][]{adam2016planckdustsed,vikram2017measurement}.
 
The tSZ effect offers a unique way to observe the diffuse baryonic gas in galaxy clusters. In those clusters, only  about 10\% of the baryons are in compact objects like stars and dust, while 90\% are in the form of diffuse gas \citep[][]{1992MNRAS.258P..14P, van2014detection}. A comparison of group and cluster masses derived from dynamical and X-ray data shows a disagreement indicating that baryons are missing in X-ray data at all scales, especially at galactic halo scales. For high-mass halos, this is likely related to the ``missing baryon'' problem occurring at redshift $z < 2$, where the intracluster gas becomes ionized in a warm phase that is particularly difficult to observe. Recently, it has also been realized that missing baryons could be a problem for the interpretation of gravitational lensing because baryonic processes could impact the dark matter distribution, even on large scales, via gravitational feedback \citep{van2011effects}. To address this issue, it is necessary to have a tracer of large scale structure probing the baryons distribution. Unlike X-ray luminosity, the tSZ signal is proportional to the baryon {pressure, which is the integral of electron number density and temperature.} This makes it easier to detect in low-density gas. 

Gravitational lensing provides an unbiased tracer of the projected mass, independent of its dynamical and physical state. cross-correlating the tSZ effect with gravitational lensing is a method to help us understand the relationship between baryons and dark matter. \citet{van2014detection} presented the first detection of a cross-correlation between the tSZ effect and weak lensing convergence with a confidence level of $6\sigma$. Later, \citet{hojjati2017cross} found a $13\sigma$ cross-correlation signal which has been used to trace the spatial distribution of the baryons relative to mass and to constrain the feedback mechanism of AGN in host galaxies. Additional constraints may be obtained by cross-correlating the tSZ signal with the CMB lensing potential, $\phi$. This signal uniquely probes the physics of intracluster gas in high-redshift, low-mass groups and clusters. \citet{hill2014detection} presented the first detection of tSZ and CMB lensing cross-correlation. They use it to constrain $\sigma_8(\Omega_m/0.282)^{0.26}$ and the intracluster medium (ICM) of galaxy clusters.

Residual systematic errors in the tSZ map may contaminate the cross-correlation results. The thermal Galactic dust emission and Cosmic Infrared Background (CIB) are two potential sources of contamination. The CIB \citep{2001ARA&A..39..249H} is the redshifted thermal emission from dust grains heated by starlight from distant galaxies. \par 

In this paper we search for residual galactic dust and CIB contamination in tSZ maps by constructing a variety of new $y$ maps with predictably different contamination levels, and by cross-correlating them with $\kappa$ and $\phi$.  For $\kappa$, we use data from the Red Sequence Cluster Lensing Survey (RCSLens), and for $\phi$, we use data from the Planck Collaboration.  The structure of this paper is as follows: Section \ref{section:formalism} introduces the formalism for our study; Section \ref{section:map_method} presents the datasets and methods; Section \ref{section:results} givess our cross-correlation results; and Section \ref{section:discussions} presentss our conclusions.\par 

\section{Formalism}
\label{section:formalism}
Both the lensing convergence, $\kappa$ and the Comptionization parameter, $y$, can be modeled as the integral of the density fluctuation $\delta_\mathrm{m}(\boldsymbol{\theta},w)$ along the line-of-sight, weighted by a kernel $W^\alpha(w)$:

\begin{equation}
I^{\alpha}(\boldsymbol{\theta}) = \int^{w_\mathrm{H}}_0 \text{d}w W^\alpha(w) \delta_m(\boldsymbol{\theta},w) \ ,
\end{equation}
where $\alpha$ denotes the component we are interested in (e.g., $\kappa$, the tSZ effect, or the CIB), $w(z)$ is the radial co-moving distance, $w_\mathrm{H}$ is the co-moving distance to the horizon, and $\boldsymbol{\theta}$ is position angle in the sky.  For $\kappa$, the kernel is

\begin{equation}
W^{\kappa}(w) = \frac{3}{2} \Omega_0 \left(\frac{H_0}{c}\right)^2g(w)\frac{d_{\mathrm{A}}(w)}{a} \ ,
\end{equation}
where $d_{\mathrm{A}}(w)$ is the angular diameter distance corresponding to $w$ and $g(w)$ depends on the redshift distribution of the sources $p_S(w)$:

\begin{equation}
g(w) = \int^{w_\mathrm{H}}_w \text{d}w' p_\mathrm{S}(w')\frac{d_{\mathrm{A}}(w'-w)}{d_{\mathrm{A}}(w')} \ .
\end{equation}

For CMB lensing, $p_\mathrm{S}(w)$ is a $\delta$-function centred at the last scattering surface. The tSZ-induced temperature change at frequency $\nu$ is characterized by the Compton parameter $y$:

\begin{equation}
\frac{\Delta T_{\mathrm{tSZ}}(\boldsymbol{\theta},x)}{T_{\text{CMB}}} = y(\boldsymbol{\theta}) S_{\text{tSZ}}(x)
\label{eq:def_yparam}
\end{equation}
where $S_{\text{tSZ}}(x) = x \coth(x/2) - 4$ is the tSZ spectral function in terms of $x \equiv h\nu/k_{\mathrm{B}}T_{\text{CMB}}$. Here $h$ is Planck's constant, $k_{\mathrm{B}}$ is the Boltzmann constant, and $T_{\text{CMB}}$ is the mean temperature of the CMB. 

The Compton parameter $y$ is given by the line-of-sight integral of the electron pressure:

\begin{equation}
y(\boldsymbol{\theta}) = \int^{w_\mathrm{H}}_0 a\text{d}w \frac{k_{\mathrm{B}}\sigma_T}{m_e c^2}n_e(\boldsymbol{\theta}, w)T_e(\boldsymbol{\theta}, w)  \ ,
\end{equation}
where $\sigma_T$ is the Thomson cross section and $n_e(\boldsymbol{\theta},w)$ and $T_e(\boldsymbol{\theta}, w)$ are the number density and temperatures of the electrons, respectively. The electron number density depends both on angular position and radial distance, $n_e(\boldsymbol{\theta},w) = \bar{n}_e \delta_{\text{gas}}(\boldsymbol{\theta},w)$ where $\bar{n}_e$ is the mean electron number density, which is proportional to $(1+z)^3$, and $\delta_{\mathrm{gas}}$ is the gas mass density fluctuation, which is given by $b_{\text{gas}}(z)\delta_m$ with $b_{\text{gas}}\propto (1+z)^{-1}$ the gas bias \citep{goldberg1999microwave}. The electron temperature depends only on radial distance, $T_e(w) \propto (1+z)^{-1}$. So the tSZ kernel is given by:

\begin{equation}
W^{\text{tSZ}}(w) = b_{\text{gas}}(0)\bar{n}_e\sigma_T \frac{k_{\mathrm{B}} T_e(0)}{m_e c^2} \frac{1}{1+z(w)}
\label{eq:wsz_df} \ .
\end{equation}

The $\kappa\times   y$ cross-correlation can be written as \citep{cooray2000large}:

\begin{equation}
C^{\kappa\times   y}_{\ell} = \int^{w_\mathrm{H}}_0 \text{d}w \left[\frac{W^{\text{tSZ}}(w)W^{\kappa}(w)}{f^2_K(w)}\right] P_m\left(\frac{\ell}{d_{\mathrm{A}}(w)},w\right) \ ,
\label{eq:cky_df}
\end{equation}
where $P_m$ is the matter power spectrum:

\begin{equation}
\langle \delta_m(\boldsymbol{k},z)\delta_m(\boldsymbol{k'},z) \rangle = (2\pi)^3 \delta(\boldsymbol{k}-\boldsymbol{k'})P_m(k,z)
\end{equation}

If we take the integral over $w$ from 0 to the last scattering surface (corresponding to $z\simeq 1100$), then the cross-correlation is between $y$ and the CMB lensing. In our analysis of $y\ \times$ CMB lensing, we use the lensing potential $\phi$ instead of $\kappa$. Note that the lensing convergence is given by $\kappa(\boldsymbol{\theta})=-\nabla_{\boldsymbol{\theta}}^2\phi(\boldsymbol{\theta})/2$ (where $\boldsymbol{\theta}$ is a unit vector pointing to the sky and $\nabla_{\boldsymbol{\theta}}^2$ is the two-dimensional Laplacian on the sky), or $\kappa_{\ell} = \ell(\ell+1)\phi_{\ell} / 2$ in multipole space.

The $y\times $CIB signal is generally described by halo model. \citet{10.1111/j.1365-2966.2012.21664.x} gives a detailed discussion about it.

\section{Data and Method}

\label{section:map_method}

\begin{table}[]
\centering
\begin{tabular}{llll}
\hline\hline
\pbox{20cm}{Frequency\\{[}GHz{]}} & \pbox{20cm}{Resolution\\{[}arcmin{]}} & Units                \\
\hline
100 & 9.66  & $\mu\mathrm{K_{CMB}}$ \\
143  & 7.27   & $\mu\mathrm{K_{CMB}}$ \\
217   & 5.01  & $\mu\mathrm{K_{CMB}}$ \\
353   & 4.86  & $\mu\mathrm{K_{CMB}}$   \\
545   & 4.84  & MJy / sr   \\
857  & 4.63   & MJy / sr \\
\hline\hline               
\end{tabular}
\caption{Overview of Planck HFI maps. Columns shows FWHM of beams, and units of raw sky maps.}
\label{table:HFI-maps}
\end{table}

Our analysis of the CIB and Galactic dust contamination is based on the cross-correlations between the different components. The maps we use are all in \texttt{\texttt{HealPix}} format with $\mathrm{N_{side}}=2048$, 1.7 arcmin pixels. 

We reconstruct a $y$  map from 6 Planck 'full mission' HFI all-sky temperature maps at 100, 143, 217, 353, 545, and 857 GHz, from Planck's 2nd data release \citep{adam2016planckmap}. Information about these band maps is given in Table \ref{table:HFI-maps}. The mask associated with the reconstructed $y$ map is a union of the 40\% galactic mask and point source mask. We also include a CIB mask from zero-signal pixels in the Planck CIB maps (see below). The joint mask excludes 47.21\% of the sky. The details of our $y$ map reconstruction are given in Appendix \ref{app:recy}. The reconstructed $y$ map is shown in Fig.\ref{fig:recymap}. We also use the Planck NILC $y$ map for comparison.

We use the Planck CIB maps \citep{aghanim2016planckcib} to help evaluate the contamination. We use the CIB maps cover the 3 highest frequencies: 353GHz, 545GHz and 857GHz, which cover about 40\%  of the sky near the galactic plane masked out. They are made by subtracting CIB-free galactic dust maps from CIB-contaminated dust maps. Both kinds of maps are constructed with a Generalized NILC method, but with different frequency-frequency covariance matrices. We also make model CIB maps at 100-217 GHz by scaling the 353 GHz CIB map to these frequencies using the grey-body CIB spectrum \eqref{eq:cibspectrum}.  However, this spectrum is redshift-dependent and we adopt a model evaluated at $z=1.2$. Details are given in Appendix \ref{subsec:cib-sub}. All the CIB maps have an angular resolution of 5 arcmin. We also apply the 40\% galactic mask to the CIB maps.

Thermal dust radiation can be modeled as a greybody spectrum with a dust spectral index $\beta_\text{d}$ and dust temperature $T_\text{d}$ \citep{hildebrand1983determination}. Both parameters are spatially dependent and were mapped by Planck collaboration \citep{ade2016planck}. We adopt the Planck dust model maps to test the robustness of the cross-correlation signal to the dust model used to null the thermal dust component.  \par

\begin{figure}\centering

\includegraphics[width=\linewidth]{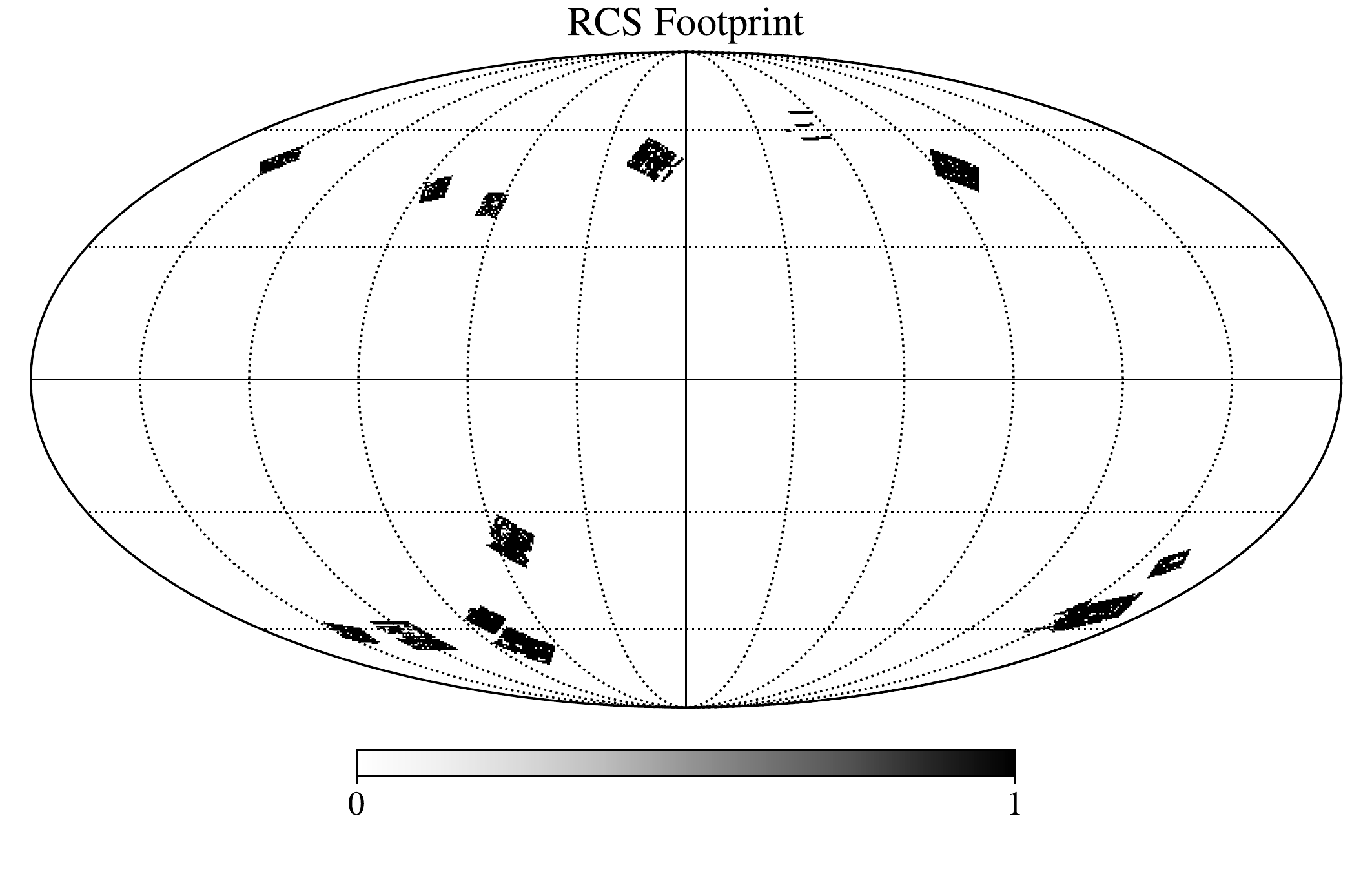}

\caption{ Footprint of RCSLenS field in galactic coordinate.}
\label{fig:rcsfield}
\end{figure}

\begin{figure}\centering

\includegraphics[width=\linewidth]{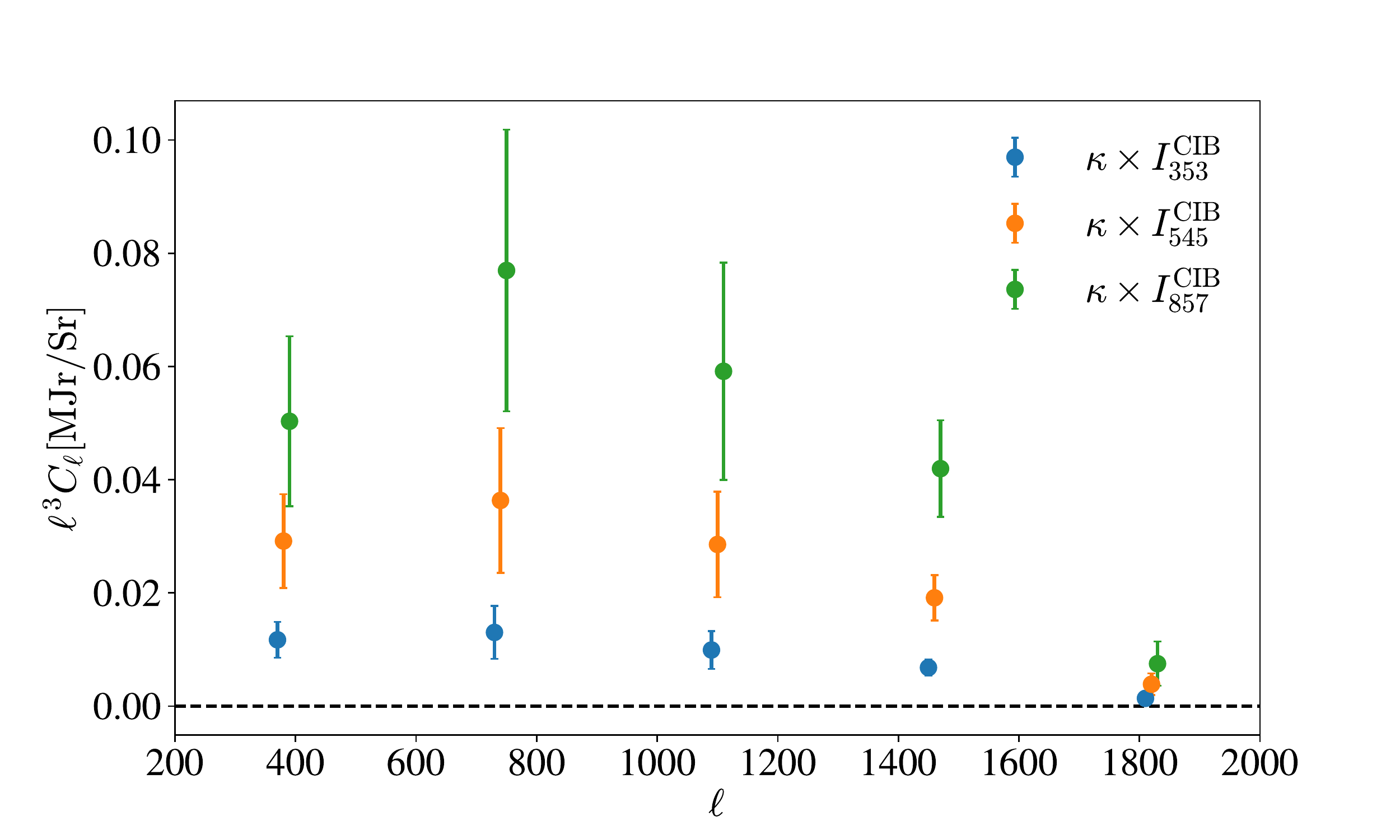}

\caption{$\kappa\times \mathrm{CIB}$ at three different frequencies in harmonic space. The cross-correlation signal is binned into 5 $\ell$ bins centered at {290,  670, 1050, 1430, 1810}. }
\label{fig:kappacib}
\end{figure}

\begin{figure} \centering

\label{fig:ky_cib}
\includegraphics[width=\columnwidth]{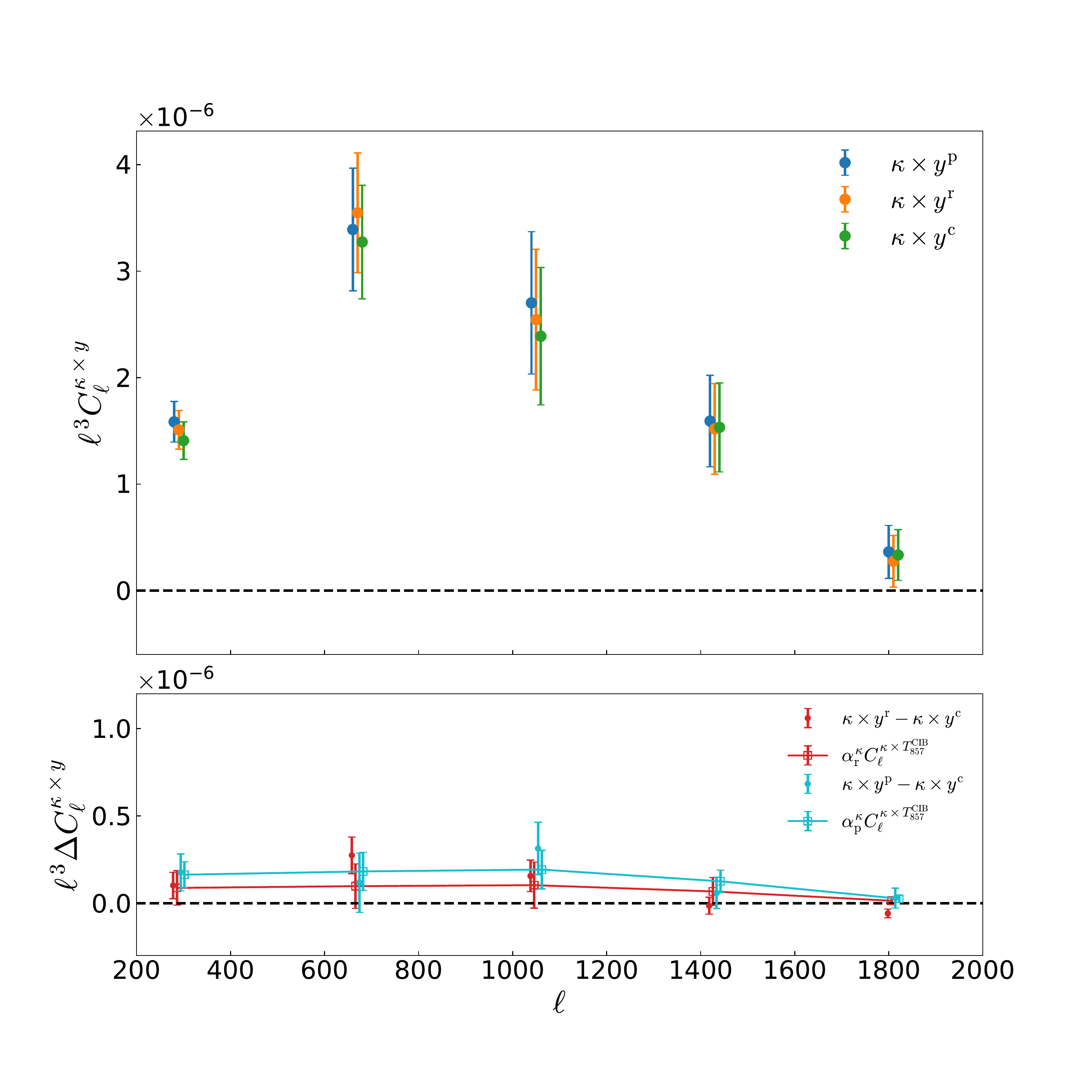}

\caption{Upper panel: $\kappa\times y$ for three $y$ maps in $\ell$ space. The cross-correlation signal is binned to 5 $\ell$ bins. Black, red, and green points are corresponding to Planck NILC $y$ map, our $\hat{y}^{\mathrm{rec}}$  map, our CIB\mbox{-}subtracted $y$ map (see \eqref{eq:ycibsubtracted}). Lower panel: Difference between $\kappa\times y^{\mathrm{r}}$ and $\kappa\times y^{\mathrm{c}}$ (blue points with errorbars), and between $\kappa\times y^{\mathrm{p}}$ and $\kappa\times y^{\mathrm{c}}$ (orange points with errorbars). Lines show the $\kappa\times T^{\mathrm{CIB}}_{857}$ scaled by the best-fit $\alpha^{\kappa}$.}

\label{fig:kappay}
\end{figure}

The galaxy lensing data is from Red Cluster Sequence Lensing Survey (RCSLenS) which is part of the second Red Sequence Cluster Survey \citep{hildebrandt2016rcslens}. Data was acquired with the MegaCAM camera from 14 separate fields and covers a total area of 785 deg$^2$ of the sky. The RCSLenS lensing signal has been cross-correlated with Planck Lensing \citep{harnois2016cfhtlens} in order to probe AGN feedback \citep{hojjati2017cross}. For our analysis we use the reconstructed projected mass map, $\kappa$. The lensing reconstruction method is developed by \citet{van2013cfhtlens}, and the map is converted to a \texttt{\texttt{HealPix}} map with $\mathrm{N_{side}}=2048$. The corresponding mask is the RCSLenS footprint.

The CMB lensing data is from the publically released Planck CMB 2018 lensing potential map \citep{aghanim2018planck}. It is constructed by a minimum-variance lensing reconstruction over 70\% of the sky. We use the tSZ deprojected lensing map, which is produced using filtered temperature data from the SMICA DX12 tSZ-deprojected CMB map. The CMB lensing data is provided in spherical harmonics of convergence $\kappa_{\ell m} = \ell(\ell+1)\phi_{\ell m}/2$, in $0<\ell<2048$. We reconstruct the $\phi$ map by inverse spherical harmonic transformation. The corresponding mask is also given in the same file package. We also analyse the effect of tSZ residual in the CMB lensing map, see Appendix \ref{app:sys_cmblens}.

We use the \texttt{PolSpice} package \cite[][]{challinor2011polspice} to calculate cross-correlation functions. We do not deconvolve the beam since all the maps have the same angular resolution. The cross-correlation  are binned into 5 $\ell$ bins centered at $\ell = \{290,  670, 1050, 1430, 1810 \}$. 

{The statistical cross-correlation uncertainties between signal A and B are assumed to be Gaussian errors which are calculated using the measured auto-spectra of signal A and B.}

\begin{equation}
\left( \Delta C _ { \ell \mathrm{, Gaussian}} ^ { \mathrm{AB}} \right) ^ { 2 } = \frac { 1 } { f _ { \mathrm { sky }} } \frac { 1 } { ( 2 \ell + 1 ) \Delta \ell } \left[ C ^ { \mathrm{AA} } C _ { \ell } ^ { \mathrm{BB} } + \left( \tilde{C} _ { \ell } ^ { \mathrm{AB} } \right) ^ { 2 } \right] \ ,
\end{equation}
where $C _ { \ell } ^ { \mathrm{AA}}$ and $C _ { \ell } ^ { \mathrm{BB}}$ are the measured auto-spectra of A and B signal; $C _ { \ell } ^ { \mathrm{AB}}$ is the measured cross-spectra. and $\tilde{C} _ { \ell } ^ { \mathrm{AB}}$ is the theoretical cross-spectrum which is much less than the auto-spectra so is negligible \citep{ade2014planckglcib}. $\Delta \ell$ is the bin width, and $f_{\mathrm{sky}}$ is the sky fraction over which the angular spectra are measured. We find that these errors are slightly lower than the measured scatter within an $\ell$-bin. In the following results, we adopt the same $\ell$ bins and error recipe throughout. This method is also used by \citet{ade2014planckglcib} and \citet{hill2014detection}. 

{
For cross-correlations involving the $y$ map, we also take the calibration uncertainty \citep{adam2016planck_inst} into account.To estimate this uncertainty, we first sample 20 sets of frequency band maps by multiplying the original sky maps with Gaussian-distributed random numbers centered at 1 with standard deviation given in \citep{adam2016planck_inst}, and then make 20 samples of $y$ maps from them. The calibration uncertainty that propagates into $C_{\ell}^{Ay}$ are $C_{\ell}^{A\sigma_y}$ where $\sigma_y$ is the standard deviation of these $y$ map samples. We assume that this uncertainty is uncorrelated with the statistical uncertainty discussed above, so the overall cross-correlation uncertainty is}
\begin{equation}
    {\left( \Delta C _ { \ell } ^ { \mathrm{A}y} \right) ^ { 2 } = \left( \Delta C _ { \ell \mathrm{, Gaussian}} ^ { \mathrm{A}y} \right) ^ { 2 } + \left( C _ { \ell} ^ { \mathrm{A}\sigma_y} \right) ^ { 2 }}
\end{equation}

\section{Results}
\label{section:results}
\subsection{CIB contamination in $\kappa\times y$ and $\phi \times y$}

\begin{figure} \centering

\subfigure { \label{fig:yp_cib}
\includegraphics[width=\columnwidth]{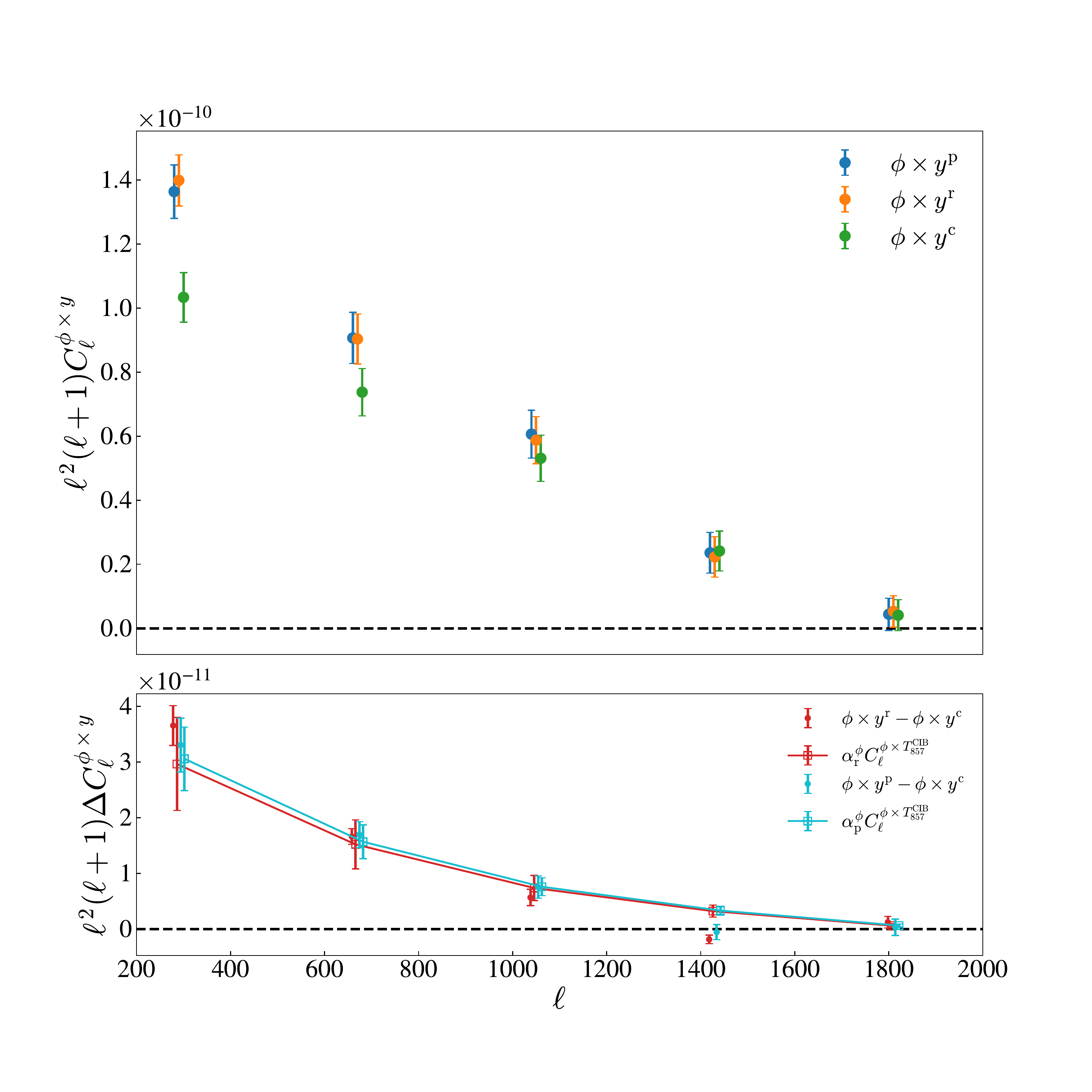}

}

\caption{Upper panel: $\phi\times y$ for three $y$ maps in $\ell$ space. Lower panel: Difference between $\phi\times y^{\mathrm{r}}$ and $\phi\times y^{\mathrm{c}}$, and between $\phi\times y^{\mathrm{p}}$ and $\phi\times y^{\mathrm{c}}$. Lines show the $\phi\times T^{\mathrm{CIB}}_{857}$ scaled by the best-fit $\alpha^{\phi}$. The colors are the same as Fig \ref{fig:ky_cib}}

\label{fig:phiy}
\end{figure}

Since the CIB traces the spatial distribution of distant galaxy clusters, it should have a non-zero cross-correlation with gravitational lensing by large scale structure. We first estimate the cross-correlation between the 3 Planck CIB maps and the RCSLens $\kappa$ map. The results are shown in Fig \ref{fig:kappacib}.  All the three CIB maps show a non-zero cross-correlation signal with a confidence level of $\sim7\sigma$. This significance is derived from the p-value of the null test given by the $\chi^2$ statistics. In this section we test the potential CIB contamination in our reconstructed $y$ map.

We denote the reconstructed $y$ map as $y^{\mathrm{r}}$ and CIB-cleaned $y$ map as $y^{\mathrm{c}}$. We parametrize CIB contamination in the L$\times y$ cross-correlation by approximating as a scaling of L$\times T^{\mathrm{CIB}}_{857}$, where $T^{\mathrm{CIB}}_{857}$ is the CIB temperature at 857 GHz and "L" denotes for lensing signal, either $\kappa$ or $\phi$:

\begin{table}[]
\centering
\label{table:cibcont}
\begin{tabular}{lll}
\hline\hline
                                                & Planck NILC  & Reconstructed \\ \hline
$\alpha^{\kappa}(\times10^{-6}\mathrm{K}^{-1})$ & \alphakcnilcn & \alphakcn      \\
$\alpha^{\phi}(\times10^{-6}\mathrm{K}^{-1})$   & \alphapcnilcn & \alphapcn      \\
$\Delta_{\kappa\times y}(\%)$                   & \kcbiasnilcn  & \kcbiasn       \\
$\Delta_{\phi\times y}(\%)$                     & \pcbiasnilcn  & \pcbiasn       \\ \hline\hline
\end{tabular}
\caption{CIB contamination in Lensing $\times y$ signal. $\alpha$ and $\Delta$ are defined in \eqref{eq:param_cibcont} and \eqref{eq:cibctfrac}}
\end{table}

\begin{equation}
\label{eq:param_cibcont}
C_{\ell}^{\mathrm{L}\times y^{\textrm{r}}} = C_{\ell}^{\mathrm{L}\times y^{\mathrm{c}}} + \alpha^{\mathrm{L}} C_{\ell}^{\mathrm{L}\times T^\mathrm{CIB}_{857}} \ ,
\end{equation}
where $\alpha^{\mathrm{L}}$ is the scaling parameter quantifying the mean CIB contamination in L$\times y$ signal \citep{hill2014detection}. Note that $\alpha$ has units of K$^{-1}$. We seek to measure $\alpha^{\kappa}$ and $\alpha^{\phi}$ from $\kappa\times y$ and $\phi\times y$.

\citet{ade2014planckglcib} showed that $\kappa\times T^\mathrm{CIB}_{857}$ and $\phi\times T^\mathrm{CIB}_{857}$ are sourced by galaxies at different redshifts ranging out to $z\sim 5$ . The RCSLenS sources ranges from 0 to $z\sim1.5$ with a peak at $z\sim 0.5$, so $\kappa\times T^\mathrm{CIB}_{857}$ is mainly from nearby galaxies and galaxy clusters. $\phi\times T^\mathrm{CIB}_{857}$ is sourced by all the CIB sources out to $z\sim 5$. Given that the CIB contamination in the $y$ map could vary with redshift, $\alpha^{\kappa}$, and $\alpha^{\phi}$ might not have the same value.

First, we measure $C_{\ell}^{\mathrm{L}\times y^{\mathrm{r}}}$,  $C_{\ell}^{\mathrm{L}\times y^{\mathrm{c}}}$, and $C_{\ell}^{\mathrm{L}\times T^\mathrm{CIB}_{857}}$, then constrain $\alpha^{\kappa}$ and $\alpha^{\phi}$, using both the Planck NILC $y$ map and our reconstructed $y$ map. The fractional L contribution in the cross-correlation signal is given by:

\begin{equation}
\label{eq:cibctfrac}
\Delta_{\mathrm{L}\times \mathrm{CIB}} = \mathrm{avg}\left(\frac{C_{\ell}^{\mathrm{L}\times y^{\textrm{r}}}-C_{\ell}^{\mathrm{L}\times y^{\mathrm{c}}}}{C_{\ell}^{\mathrm{L}\times y^{\textrm{r}}}}\right) \ .
\end{equation}

We perform a same analysis for the Planck NILC $y$ map $y^{\textrm{p}}$.

The $\kappa\times y$ results are shown in Fig \ref{fig:kappay}. The three sets of points with 3 different colors correspond to 3 different  $y$  maps correlated with a common $\kappa$ map: the Planck NILC $y$ map $y^{\textrm{p}}$, our reconstructed $y$ map $y^{\textrm{r}}$, and our CIB\mbox{-}subtracted $y$ map $y^{\textrm{c}}$. All 3 cross-correlations show a nonzero $\kappa\times y$ signal with  $> 5\sigma$ confidence level. $y^{\textrm{r}}$ has an 8$\sigma$ signal which is consistent with \citet{hojjati2016cross}. Note that $\kappa\times y^{\textrm{r}}$ agrees with $\kappa\times y^{\textrm{p}}$ at $1\sigma$ level. The $\phi\times y$ signal is analysed in the same way as $\kappa\times y$. The results are presented in Fig.\ref{fig:phiy}. The signal is more significant than $\kappa\times y$ because the sky fraction covered by CMB lensing is much larger than RCSLenS.

A summary of CIB contamination in L$\times y$ estimation is given in Table \ref{table:cibcont}. The CIB contamination in $\kappa\times y$ has a significance of $\sim 2\sigma$ while it has a significance of $\sim 5\sigma$ in $\phi\times y$ for both $y^{\mathrm{r}}$ and $y^{\mathrm{p}}$. {\citet{hill2014detection} makes a similar detection by subtracting a CIB bias directly from the $y\times \phi$ cross-correlation.} \citet{hurier2015predicting} gets a $(20\pm 10)\%$ $\phi\times\mathrm{CIB}$ contamination in their $\phi\times y$ measurement which is similar to our measurement. However, our measurement is more accurate due to the larger sky coverage. From the estimated $\alpha$ values we conclude that the CIB contamination in the Planck NILC $y$ map and the reconstructed $y$ map is at a level of $\sim10^{-6}(T^{\mathrm{CIB}}_{857}/\mathrm{K})$ or $\sim10^{-1}(T_{\mathrm{CIB143}}/\mathrm{K})$. 

When constructing the CIB-cleaned $y$ map, a CIB model is used to generate the model CIB maps at 100, 143, 217 GHz (see section \ref{subsec:cib-sub} for details). Our results assume that the Planck CIB maps are not contaminated by tSZ signal. The {\em rms} signal levels in the three Planck CIB maps are 1.03, 2.72, and 5.03 $\times 10^{-2}$ MJy/sr at 353, 545, and 857 GHz, respectively.  The {\em rms} of the estimated tSZ signal at these frequencies is 0.76, 0.36, and 0.03 $\times 10^{-2}$ MJy/sr.  Thus the CIB map at 353 GHz has a relatively high risk of being contaminated by uncorrected tSZ signal.   To test this assumption, we have performed an alternate scaling of the CIB signal using the 545 GHz CIB map as a template, and scaling it to 100-353 GHz bands using the same spectral model as above, evaluated at $z=1.2$. We make another CIB-cleaned $y$ map with them and do a similar cross-correlation analysis. We find the contamination fractions are lower by about $\sim 1\sigma$ (see Table \ref{table:cibcont545}). It indicates that the tSZ residual in the 353GHz CIB map does not affect our contamination estimation significantly.

We have also tested robustness to the assumed CIB model redshift. We find that when the CIB redshift varies from 0.8 to 2.0, the cross-correlation results change by $<1\sigma$, so our results are not sensitive to CIB redshift. For the 545-scaled CIB maps, the results are somewhat more sensitive to model redshift.

\begin{table}
\centering
\label{table:cibcont545}
\begin{tabular}{lll}
\hline\hline
                                                & Planck NILC  & Reconstructed \\ \hline
$\alpha^{\kappa}(\times10^{-6}\mathrm{K}^{-1})$ & \alphakcnilcnp & \alphakcnp      \\
$\alpha^{\phi}(\times10^{-6}\mathrm{K}^{-1})$   & \alphapcnilcnp & \alphapcnp      \\
$\Delta_{\kappa\times y}(\%)$                   & \kcbiasnilcnp  & \kcbiasnp       \\
$\Delta_{\phi\times y}(\%)$                     & \pcbiasnilcnp  & \pcbiasnp       \\ \hline\hline
\end{tabular}
\caption{Same table as Table.\ref{table:cibcont}, but with different CIB-cleaned $y$ map with 100-353GHz CIB map extrapolated from 545GHz CIB map.}
\end{table}

\subsection{Robustness to Galactic Dust Contamination}

The goal of this section is to evaluate the robustness of tSZ-lensing cross-correlations if a galactic dust-spectrum is projected-out when making $y$ maps (section \ref{subsec:dust-nulled} discusses how these maps are made). We vary the dust spectral index $\beta_\mathrm{d} $ to project out different models. Since galactic dust cannot be described by a single spectrum, these projection procedures surely leaves different level of dust residual in the $y$ maps.

In this investigation, CIB is removed beforehand to control the CIB residual in the $y$ maps. Since the CIB maps are contaminated by some level of galactic dust, this subtraction leads to additional influence on dust residual in the $y$ maps.

\begin{figure}\centering

 \label{fig:ky_dustvalue}
 
\includegraphics[width=\columnwidth]{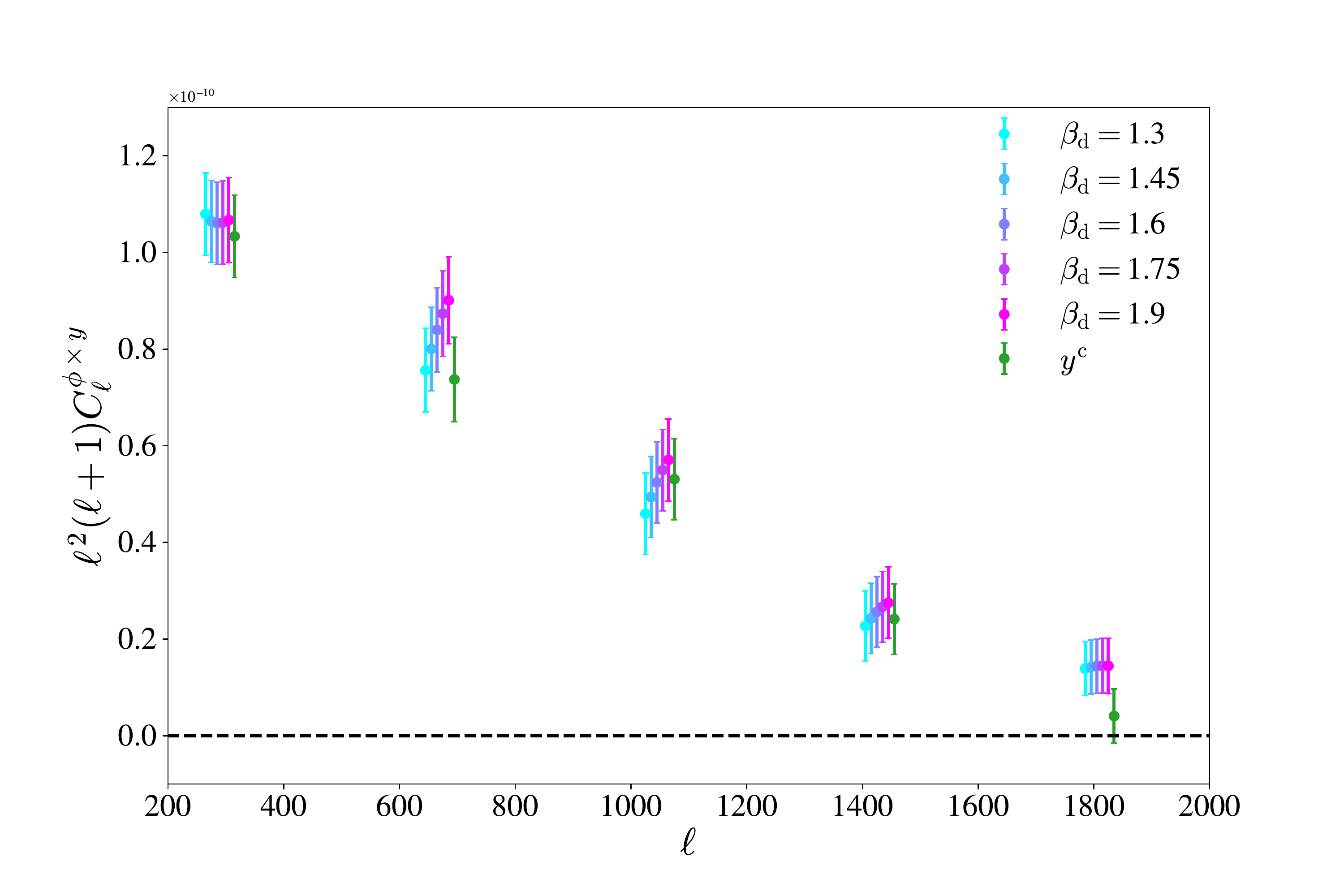}

\caption{ $\phi\times  y$ cross-correlation measured from different dust-nulled $y$ maps. The $\ell$ bins are the same as Fig \ref{fig:ky_cib}. $\kappa\times y^{\mathrm{c}}$ is also plotted as a reference using green points. Note that $y^{\mathrm{c}}$ is with no dust model corrected.}
\label{fig:py_dust}
\end{figure}

Both $\kappa\times y$ (Fig \ref{fig:ky_dust}) and $\phi\times y$ (Fig \ref{fig:py_dust}) show that the cross signals don't change when we vary the Galactic dust models (see Fig.\ref{fig:dustmodel}). This is consistent with our expectation that Galactic dust is not correlated with extragalactic signals.

\section{Discussion and Conclusion}
\label{section:discussions}

\begin{figure}\centering

 \label{fig:ky_dustvalue}
\includegraphics[width=\columnwidth]{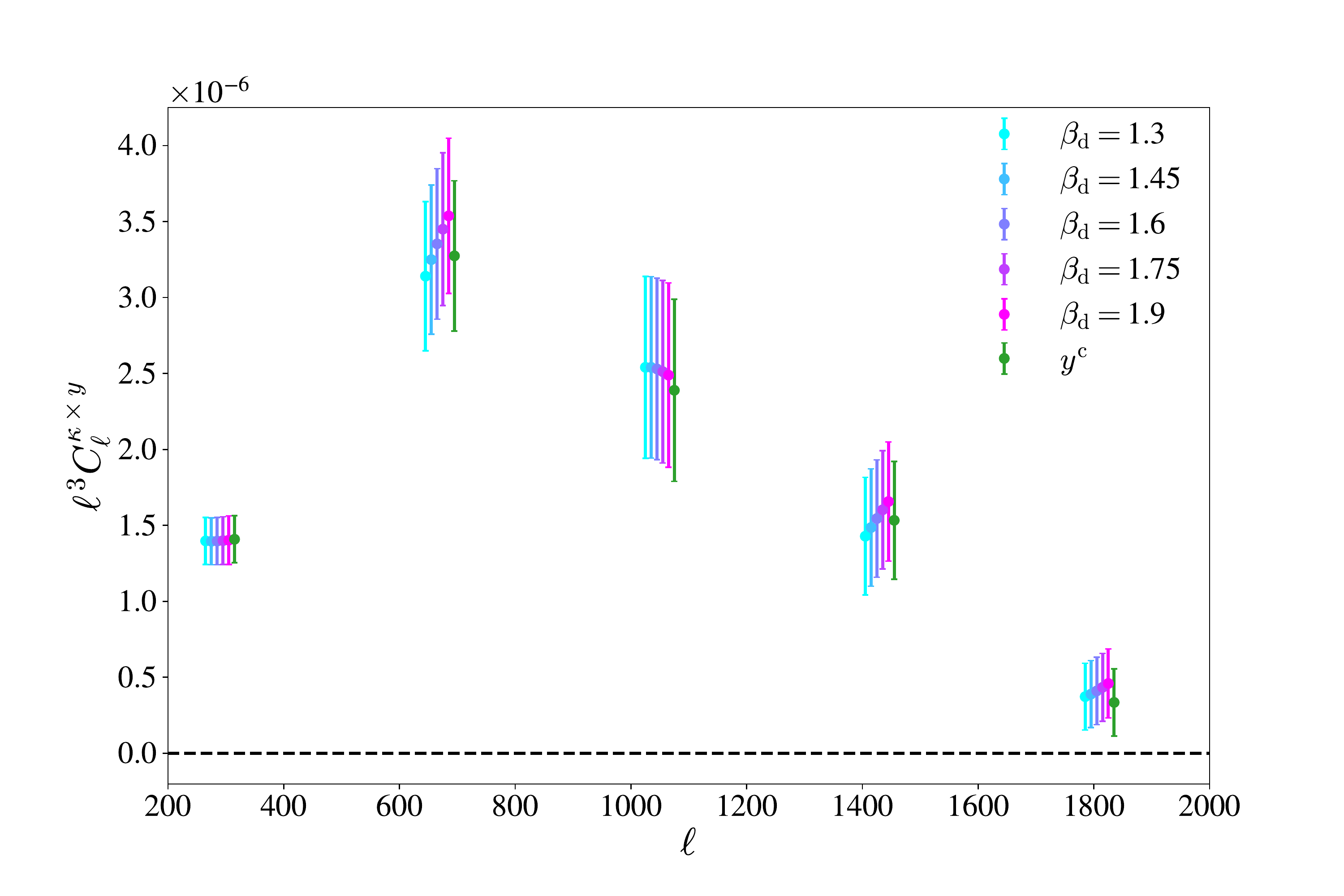}

\caption{ $\kappa\times  y$ cross-correlation measured from different dust-nulled $y$ maps. The $\ell$ bins are the same as Fig \ref{fig:ky_cib}. $\kappa\times y^{\mathrm{c}}$ is also plotted as a reference using green points.  Note that $y^{\mathrm{c}}$ is with no dust model corrected.}
\label{fig:ky_dust}
\end{figure}

We have estimated the degree to which emission from CIB and Galactic dust contaminates a set of tSZ maps, $y$, and their cross-correlation with gravitational lensing ($\kappa$ or $\phi$).  We do so by modifying the procedure for reconstructing $y$ maps from the Planck band maps.  We first verify that we can reconstruct the nominal NILC map produced by the Planck team, denoted here as $y$, up to small differences detailed in Appendix A.   We denote our nominal reconstruction $y^{\mathrm{r}}$.  We then produce a modified tSZ map by subtracting a CIB model from each band map prior to forming a tSZ map.  We denote this corrected map $y^{\mathrm{c}}$ and we recompute the lensing cross-correlation signal with this map.  Under the assumption that ${\kappa \times y^{\mathrm{c}}}$ is not contaminated by CIB emission, we estimate that the cross-correlation with the original Planck tSZ map, $\kappa \times y^{\mathrm{p}}$, is contaminated by \kcbiasnilc, and the cross-correlation with our nominally-reconstructed tSZ map, $\kappa \times y^{\mathrm{r}}$, is contaminated by \kcbias.  In neither case is the contamination significant.

\citet{van2014detection} presented a first detection of $\kappa \times y$ and used the measurement to constrain $b_{\rm gas} \, (T_e/1 \, {\rm keV}) \, (n_e/1 \, {\rm cm}^{-3}) = 2.01 \pm 0.31 \pm 0.21$, where $b_{\rm gas}$ is the gas bias, $T_e$ is the mean electron temperature (at redshift zero), and $n_e$ is the mean electron number density.  The first uncertainty is statistical and the second is an estimate of the systematic error.  Our work shows that this result is unlikely to be significantly affected by CIB contamination. However, as future lensing surveys increase their sky coverage, CIB contamination will likely become a statistically significant source of bias in measurements of ${\kappa \times y}$.

We perform a similar cross-correlation analysis with the CMB lensing potential, $\phi \times y$, and find a significant contamination: \pcbiasnilc in the nominal Planck tSZ map, and \pcbias in our nominally-reconstructed map.  We attribute this higher level of contamination to the fact that the redshift distribution of CMB lensing sources overlaps more with that of the CIB sources than do the galaxy lensing sources. And also $\phi$ signal covers a larger fraction of the sky, {which reduces the uncertainty in cross-correlation signal and thus enhances the significance of the CIB contamination.}.

Since the CIB redshift is fixed at 1.2 when extrapolating low frequency CIB maps, CIB emission from sources at other redshifts are not completely removed.  We test the robustness of our extrapolation by evaluating the spectral model at other redshifts in the range $0.8 < z < 2.0$ and find that our cross-correlation bias results are stable.

We test the effect of tSZ residuals in the 353GHz CIB map and find that the results do not change significantly when we use the CIB-cleaned map with 100-353GHz CIB model extrapolated from 545GHz CIB map. However, we also find that the latter is more sensitive to the CIB redshift we choose. \par  

Emission from Galactic dust is uncorrelated with the tSZ and CIB emission and it does not contaminates our results significantly.  We vary the residual dust signal in our reconstructed $y$ maps by varying the dust spectral index $\beta_d$ when we produce them.  The cross-correlation results in Fig.\ref{fig:ky_dust} show that $\kappa \times y^{\mathrm{r}}$ is not sensitive to  the dust model. \par 

Ongoing and future weak lensing observations will continue to expand their sky coverage and redshift range.  We anticipate that it will soon be feasible to separate CIB and tSZ signals in the far-infrared data with high precision, allowing for very robust measurements of $\kappa \times y$ and $\kappa \times$ CIB.  These data will enable new insights about the evolution of gas in the universe.

\section*{Acknowledgement} We thank Dr. Hideki Tanimura for useful discussion. We also acknowledge Dr. Mathieu Remazeilles for noting us about the tSZ residual in CMB lensing map. We thank the Planck Collaboration for making its data available to download at \url{https://www.cosmos.esa.int/web/planck/pla}. This work is financially supported by Canada’s NSERC and CIFAR. TT acknowledges funding from the European Union's Horizon 2020 research and innovation programme under the Marie Sk{l}odowska-Curie grant agreement No 797794.
%
%

\bibliographystyle{apj}
\bibliography{reference}

\begin{thebibliography}{}
\expandafter\ifx\csname natexlab\endcsname\relax\def\natexlab#1{#1}\fi

\bibitem[{Adam {et~al.}(2016{\natexlab{a}})Adam, Ade, Aghanim, Arnaud, Ashdown,
  Aumont, Baccigalupi, Banday, Barreiro, Bartolo, {et~al.}}]{adam2016planckmap}
Adam, R., Ade, P., Aghanim, N., {et~al.} 2016{\natexlab{a}}, Astronomy \&
  Astrophysics, 594, A8

\bibitem[{Adam {et~al.}(2016{\natexlab{b}})Adam, Ade, Aghanim, Arnaud, Ashdown,
  Aumont, Baccigalupi, Banday, Barreiro, Bartolo,
  {et~al.}}]{adam2016planck_inst}
---. 2016{\natexlab{b}}, Astronomy \& Astrophysics, 594, A8

\bibitem[{Adam {et~al.}(2016{\natexlab{c}})Adam, Ade, Aghanim, Alves, Arnaud,
  Ashdown, Aumont, Baccigalupi, Banday, Barreiro,
  {et~al.}}]{adam2016planckdust}
---. 2016{\natexlab{c}}, Astronomy \& Astrophysics, 594, A10

\bibitem[{Adam {et~al.}(2016{\natexlab{d}})Adam, Ade, Aghanim, Ashdown, Aumont,
  Baccigalupi, Banday, Barreiro, Bartolo, Battaner,
  {et~al.}}]{adam2016planckdustsed}
---. 2016{\natexlab{d}}, Astronomy \& Astrophysics, 596, A104

\bibitem[{Addison {et~al.}(2012)Addison, Dunkley, \&
  Spergel}]{10.1111/j.1365-2966.2012.21664.x}
Addison, G.~E., Dunkley, J., \& Spergel, D.~N. 2012, Monthly Notices of the
  Royal Astronomical Society, 427, 1741

\bibitem[{Ade {et~al.}(2014{\natexlab{a}})Ade, Aghanim, Armitage-Caplan,
  Arnaud, Ashdown, Atrio-Barandela, Aumont, Baccigalupi, Banday, Barreiro,
  {et~al.}}]{ade2014planckhfi}
Ade, P., Aghanim, N., Armitage-Caplan, C., {et~al.} 2014{\natexlab{a}},
  Astronomy \& Astrophysics, 571, A9

\bibitem[{Ade {et~al.}(2011)Ade, Aghanim, Arnaud, Ashdown, Aumont, Baccigalupi,
  Balbi, Banday, Barreiro, Bartlett, {et~al.}}]{ade2011planckcib}
Ade, P.~A., Aghanim, N., Arnaud, M., {et~al.} 2011, Astronomy \& Astrophysics,
  536, A18

\bibitem[{Ade {et~al.}(2014{\natexlab{b}})Ade, Aghanim, Armitage-Caplan,
  Arnaud, Ashdown, Atrio-Barandela, Aumont, Baccigalupi, Banday, Barreiro,
  {et~al.}}]{ade2014planckglcib}
Ade, P.~A., Aghanim, N., Armitage-Caplan, C., {et~al.} 2014{\natexlab{b}},
  Astronomy \& Astrophysics, 571, A18

\bibitem[{Ade {et~al.}(2016)Ade, Aghanim, Alves, Aniano, Arnaud, Ashdown,
  Aumont, Baccigalupi, Banday, Barreiro, {et~al.}}]{ade2016planck}
Ade, P.~A., Aghanim, N., Alves, M., {et~al.} 2016, Astronomy \& Astrophysics,
  586, A132

\bibitem[{Aghanim {et~al.}(2016{\natexlab{a}})Aghanim, Arnaud, Ashdown, Aumont,
  Baccigalupi, Banday, Barreiro, Bartlett, Bartolo, Battaner,
  {et~al.}}]{aghanim2016planck}
Aghanim, N., Arnaud, M., Ashdown, M., {et~al.} 2016{\natexlab{a}}, Astronomy \&
  Astrophysics, 594, A22

\bibitem[{Aghanim {et~al.}(2016{\natexlab{b}})Aghanim, Ashdown, Aumont,
  Baccigalupi, Ballardini, Banday, Barreiro, Bartolo, Basak, Benabed,
  {et~al.}}]{aghanim2016planckcib}
Aghanim, N., Ashdown, M., Aumont, J., {et~al.} 2016{\natexlab{b}}, Astronomy \&
  Astrophysics, 596, A109

\bibitem[{Aghanim {et~al.}(2018{\natexlab{a}})Aghanim, Akrami, Ashdown, Aumont,
  Baccigalupi, Ballardini, Banday, Barreiro, Bartolo, Basak,
  {et~al.}}]{aghanim2018planck_param}
Aghanim, N., Akrami, Y., Ashdown, M., {et~al.} 2018{\natexlab{a}}, arXiv
  preprint arXiv:1807.06209

\bibitem[{Aghanim {et~al.}(2018{\natexlab{b}})Aghanim, Akrami, Ashdown, Aumont,
  Baccigalupi, Ballardini, Banday, Barreiro, Bartolo, Basak,
  {et~al.}}]{aghanim2018planck}
---. 2018{\natexlab{b}}, arXiv preprint arXiv:1807.06210

\bibitem[{Basak \& Delabrouille(2012)}]{basak2012needlet}
Basak, S., \& Delabrouille, J. 2012, Monthly Notices of the Royal Astronomical
  Society, 419, 1163

\bibitem[{Birkinshaw(1999)}]{birkinshaw1999sunyaev}
Birkinshaw, M. 1999, Physics Reports, 310, 97

\bibitem[{Challinor {et~al.}(2011)Challinor, Chon, Colombi, Hivon, Prunet, \&
  Szapudi}]{challinor2011polspice}
Challinor, A., Chon, G., Colombi, S., {et~al.} 2011, Astrophysics Source Code
  Library

\bibitem[{Chen {et~al.}(2018)Chen, Remazeilles, \& Dickinson}]{chen2018impact}
Chen, T., Remazeilles, M., \& Dickinson, C. 2018, arXiv preprint
  arXiv:1803.08853

\bibitem[{Collaboration {et~al.}(2014)}]{planck2014planckbeam}
Collaboration, P., {et~al.} 2014

\bibitem[{Cooray {et~al.}(2000)Cooray, Hu, \& Tegmark}]{cooray2000large}
Cooray, A., Hu, W., \& Tegmark, M. 2000, The Astrophysical Journal, 540, 1

\bibitem[{De~Lucia \& Blaizot(2007)}]{de2007hierarchical}
De~Lucia, G., \& Blaizot, J. 2007, Monthly Notices of the Royal Astronomical
  Society, 375, 2

\bibitem[{Delabrouille {et~al.}(2009)Delabrouille, Cardoso, Le~Jeune, Betoule,
  Fay, \& Guilloux}]{delabrouille2009full}
Delabrouille, J., Cardoso, J.-F., Le~Jeune, M., {et~al.} 2009, Astronomy \&
  Astrophysics, 493, 835

\bibitem[{Fixsen(2009)}]{fixsen2009temperature}
Fixsen, D. 2009, The Astrophysical Journal, 707, 916

\bibitem[{Goldberg \& Spergel(1999)}]{goldberg1999microwave}
Goldberg, D.~M., \& Spergel, D.~N. 1999, Physical Review D, 59, 103002

\bibitem[{Harnois-D{\'e}raps {et~al.}(2016)Harnois-D{\'e}raps, Tr{\"o}ster,
  Hojjati, van Waerbeke, Asgari, Choi, Erben, Heymans, Hildebrandt, Kitching,
  {et~al.}}]{harnois2016cfhtlens}
Harnois-D{\'e}raps, J., Tr{\"o}ster, T., Hojjati, A., {et~al.} 2016, Monthly
  Notices of the Royal Astronomical Society, 460, 434

\bibitem[{Hasselfield {et~al.}(2013)Hasselfield, Hilton, Marriage, Addison,
  Barrientos, Battaglia, Battistelli, Bond, Crichton, Das,
  {et~al.}}]{hasselfield2013atacama}
Hasselfield, M., Hilton, M., Marriage, T.~A., {et~al.} 2013, Journal of
  Cosmology and Astroparticle Physics, 2013, 008

\bibitem[{{Hauser} \& {Dwek}(2001)}]{2001ARA&A..39..249H}
{Hauser}, M.~G., \& {Dwek}, E. 2001, \araa, 39, 249

\bibitem[{Hildebrand(1983)}]{hildebrand1983determination}
Hildebrand, R.~H. 1983, Quarterly Journal of the Royal Astronomical Society,
  24, 267

\bibitem[{Hildebrandt {et~al.}(2016)Hildebrandt, Choi, Heymans, Blake, Erben,
  Miller, Nakajima, van Waerbeke, Viola, Buddendiek,
  {et~al.}}]{hildebrandt2016rcslens}
Hildebrandt, H., Choi, A., Heymans, C., {et~al.} 2016, Monthly Notices of the
  Royal Astronomical Society, 463, 635

\bibitem[{Hill \& Spergel(2014)}]{hill2014detection}
Hill, J.~C., \& Spergel, D.~N. 2014, Journal of Cosmology and Astroparticle
  Physics, 2014, 030

\bibitem[{Hojjati {et~al.}(2016)Hojjati, Tr{\"o}ster, Harnois-D{\'e}raps,
  McCarthy, van Waerbeke, Choi, Erben, Heymans, Hildebrandt, Hinshaw,
  {et~al.}}]{hojjati2016cross}
Hojjati, A., Tr{\"o}ster, T., Harnois-D{\'e}raps, J., {et~al.} 2016, arXiv
  preprint arXiv:1608.07581

\bibitem[{Hojjati {et~al.}(2017)Hojjati, Tr{\"o}ster, Harnois-D{\'e}raps,
  McCarthy, van Waerbeke, Choi, Erben, Heymans, Hildebrandt, Hinshaw,
  {et~al.}}]{hojjati2017cross}
---. 2017, Monthly Notices of the Royal Astronomical Society, 471, 1565

\bibitem[{Hurier(2015)}]{hurier2015predicting}
Hurier, G. 2015, Astronomy \& Astrophysics, 575, L11

\bibitem[{Hurier {et~al.}(2013)Hurier, Mac{\'\i}as-P{\'e}rez, \&
  Hildebrandt}]{hurier2013milca}
Hurier, G., Mac{\'\i}as-P{\'e}rez, J., \& Hildebrandt, S. 2013, Astronomy \&
  Astrophysics, 558, A118

\bibitem[{Leach {et~al.}(2008)Leach, Cardoso, Baccigalupi, Barreiro, Betoule,
  Bobin, Bonaldi, Delabrouille, De~Zotti, Dickinson,
  {et~al.}}]{leach2008component}
Leach, S.~M., Cardoso, J.-F., Baccigalupi, C., {et~al.} 2008, Astronomy \&
  Astrophysics, 491, 597

\bibitem[{Madhavacheril \& Hill(2018)}]{PhysRevD.98.023534}
Madhavacheril, M.~S., \& Hill, J.~C. 2018, Phys. Rev. D, 98, 023534

\bibitem[{Molnar {et~al.}(2002)Molnar, Birkinshaw, \&
  Mushotzky}]{molnar2002constraints}
Molnar, S., Birkinshaw, M., \& Mushotzky, R. 2002, The Astrophysical Journal,
  570, 1

\bibitem[{Narcowich {et~al.}(2006)Narcowich, Petrushev, \&
  Ward}]{narcowich2006localized}
Narcowich, F.~J., Petrushev, P., \& Ward, J.~D. 2006, SIAM Journal on
  Mathematical Analysis, 38, 574

\bibitem[{Oliver {et~al.}(2010)Oliver, Frost, Farrah, Gonzalez-Solares, Shupe,
  Henriques, Roseboom, Alfonso-Luis, Babbedge, Frayer,
  {et~al.}}]{oliver2010specific}
Oliver, S., Frost, M., Farrah, D., {et~al.} 2010, Monthly Notices of the Royal
  Astronomical Society, 405, 2279

\bibitem[{{Persic} \& {Salucci}(1992)}]{1992MNRAS.258P..14P}
{Persic}, M., \& {Salucci}, P. 1992, \mnras, 258, 14P

\bibitem[{{Planck Collaboration XXIX}(2014)}]{planck2014plancksz}
{Planck Collaboration XXIX}. 2014

\bibitem[{{Planck Collaboration XXX}(2014)}]{2014A&A...571A..30P}
{Planck Collaboration XXX}. 2014, \aap, 571, A30

\bibitem[{Schmidt {et~al.}(2014)Schmidt, M{\'e}nard, Scranton, Morrison,
  Rahman, \& Hopkins}]{schmidt2014inferring}
Schmidt, S.~J., M{\'e}nard, B., Scranton, R., {et~al.} 2014, Monthly Notices of
  the Royal Astronomical Society, 446, 2696

\bibitem[{Shang {et~al.}(2012)Shang, Haiman, Knox, \& Oh}]{shang2012improved}
Shang, C., Haiman, Z., Knox, L., \& Oh, S.~P. 2012, Monthly Notices of the
  Royal Astronomical Society, 421, 2832

\bibitem[{Tanimura {et~al.}(2017)Tanimura, Hinshaw, McCarthy, Van~Waerbeke, Ma,
  Mead, Hojjati, \& Tr{\"o}ster}]{tanimura2017search}
Tanimura, H., Hinshaw, G., McCarthy, I.~G., {et~al.} 2017, arXiv preprint
  arXiv:1709.05024

\bibitem[{van Daalen {et~al.}(2011)van Daalen, Schaye, Booth, \&
  Dalla~Vecchia}]{van2011effects}
van Daalen, M.~P., Schaye, J., Booth, C., \& Dalla~Vecchia, C. 2011, Monthly
  Notices of the Royal Astronomical Society, 415, 3649

\bibitem[{Van~Waerbeke {et~al.}(2014)Van~Waerbeke, Hinshaw, \&
  Murray}]{van2014detection}
Van~Waerbeke, L., Hinshaw, G., \& Murray, N. 2014, Physical Review D, 89,
  023508

\bibitem[{Van~Waerbeke {et~al.}(2013)Van~Waerbeke, Benjamin, Erben, Heymans,
  Hildebrandt, Hoekstra, Kitching, Mellier, Miller, Coupon,
  {et~al.}}]{van2013cfhtlens}
Van~Waerbeke, L., Benjamin, J., Erben, T., {et~al.} 2013, Monthly Notices of
  the Royal Astronomical Society, 433, 3373

\bibitem[{Vikram {et~al.}(2017)Vikram, Lidz, \& Jain}]{vikram2017measurement}
Vikram, V., Lidz, A., \& Jain, B. 2017, Monthly Notices of the Royal
  Astronomical Society, stw3311

\bibitem[{Zeldovich \& Sunyaev(1969)}]{zeldovich1969interaction}
Zeldovich, Y.~B., \& Sunyaev, R. 1969, Astrophysics and Space Science, 4, 301

\end{thebibliography}

\appendix
\twocolumngrid

\section{CIB Flux Model}
\label{app:recy}
\begin{table}[]
\centering

\label{my-label}
\begin{tabular}{ll}
\hline\hline
Parameter              & Mean Value    \\
\hline
$\alpha_{\mathrm{CIB}}$ & $0.36\pm0.05$ \\
$T_{c0}$               & $24.4\pm1.9$    \\
$\beta_{\mathrm{CIB}}$  & $1.75\pm0.06$ \\
$\gamma_{\mathrm{CIB}}$ & $1.7\pm0.2$   \\
$\delta_{\mathrm{CIB}}$ & $3.6\pm0.2$  \\
\hline\hline
\end{tabular}
\label{table:cibmodel}
\caption{Best fit CIB parameters from \citet{2014A&A...571A..30P}.}
\end{table}

\begin{figure}\centering

\includegraphics[width=\linewidth]{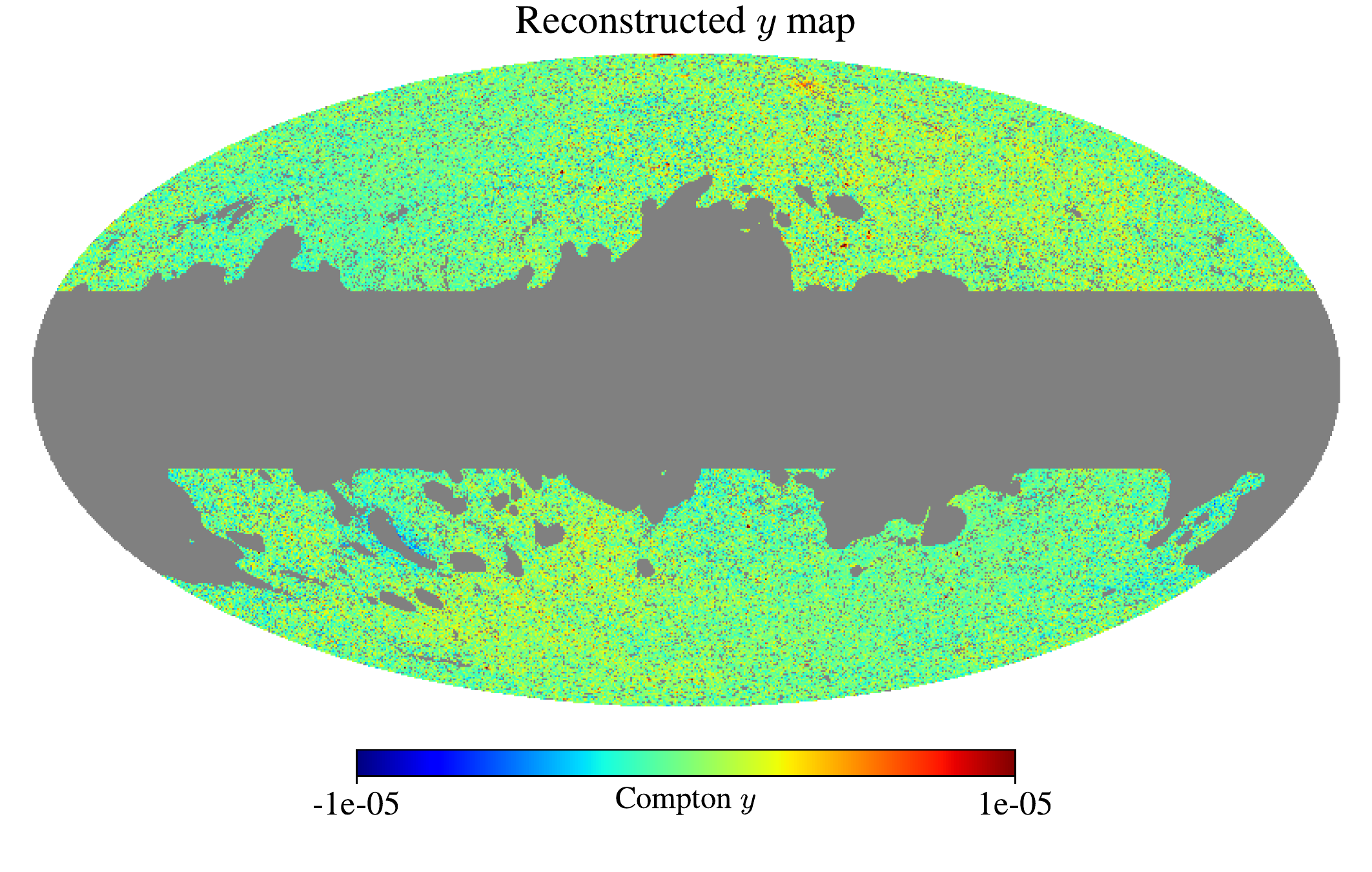}

\caption{ The fiducial $y$ map reconstructed by our {HILC} pipeline shown in Galactic coordinates.}
\label{fig:recymap}
\end{figure}

The CIB flux from a single galaxy cluster is modeled by the integrated luminosity within 500 times virial radius \citep{shang2012improved}:

\begin{equation}
L^{\text{CIB}}_{500}(\nu,z) = L_0\left[\frac{M_{500}}{10^{14}M_{\odot}}\right]^{\epsilon_{\text{CIB}}}\Psi(z)\Theta[(1+z)\nu,T_c(z)] \ ,
\label{eq:cibint}
\end{equation}
where $L_0$ is a normalization parameter, $T_\text{c} = T_\text{c0}(1+z)^{\alpha_{\text{CIB}}}$ is the dust temperature of the cluster.  $T_{\mathrm{c0}}$ is the current-time dust temperature and $\alpha_{\text{CIB}}$ accounts for the distinctive evolution of dust temperature. $\Theta[\nu,T_\text{c}]$ is the SED for a typical galaxy that contributes to the total CIB emission,

\begin{equation}
\Theta[\nu,T_\text{c}] = 
 \begin{cases}
      \nu ^{\beta_{\text{CIB}}}B_{\nu}(T_\text{c}), & \ \nu < \nu_0 \\
      \nu ^{-\gamma_{\text{CIB}}}, & \ \nu \geq \nu_0 
    \end{cases}\ .
    \label{eq:cibspectrum}
\end{equation}
Here $\mathrm{B}_{\nu}$ denotes the blackbody intensity, while  the  emissivity index $\beta$ gives information about the physical nature of dust which in general depends on grain composition. $\nu_0$ is the solution of $\text{d}\log[\nu^{\beta_{\text{CIB}}}B_{\nu}(T_\text{c})] / \text{d}\log(\nu) = -\gamma_{\text{CIB}}$, which connects the SED at high and low frequencies. The redshift dependence is assumed to be the form:

\begin{equation}
\Psi = (1+z)^{\delta_{\text{CIB}}} \ .
\end{equation}
This dependence has been justified by observations \citep[e.g., ]{de2007hierarchical, oliver2010specific}.

The model parameters we use are given in Table.\ref{table:cibmodel}. We use the frequency dependence of CIB to generate model CIB maps at 100-217GHz in order to subtract CIB from $y$ map.

\section{Reconstruction of the $y$ Signal}
\label{app:recy}

\subsection{The Internal Linear Combination}

The raw temperature maps are a superposition of different emission components, including the CMB, galactic dust emission, free-free radiation, synchrotron radiation, the CIB,  tSZ signal etc. A well-known method to extract one of those components with known frequency spectrum and null others is the Internal Linear Combination (ILC) technique, used by WMAP and Planck to make the CMB and other component maps \citep{leach2008component}.  The basic idea of the ILC method is to use a linear combination of different frequency maps to keep the target component unbiased and to minimize the contribution of other components.\par 

In our analysis, in addition to minimize other contamination, we also include a constrain to project the CMB out. The CMB and tSZ signals are separable, $s^{\alpha}_{\nu}=f^{\alpha}(\nu)s_{\alpha}(\boldsymbol{\theta})$, where $f^{\alpha}(\nu)$ is the frequency dependence for component $\alpha$ ($\alpha=$CMB, tSZ) at frequency $\nu$. $s_{\alpha}(\boldsymbol{\theta})$ is the sky template of component $\alpha$. The objective of ILC is to extract $s_{\alpha}(\boldsymbol{\theta})$.

For a given detector, the frequency dependence is weighted by the bandpass:

\begin{equation}
f^{\alpha}_i = \frac{\int b_i(\nu)f^{\alpha}(\nu)\mathrm{d}\nu}{\int b_i(\nu)\mathrm{d}\nu} \ ,
\end{equation}
where $b_i(\nu)$ is the bandpass of the detector whose frequency response is centered at $\nu_i$. For Planck satellite, they are given in \citet{ade2014planckhfi}. We use Latin letters $i,j,k...$ for frequency channels and Greek letters $\alpha,\beta,\gamma ...$ for components hereafter.

The raw sky map at frequency $\nu_i$ is a combination of all the components plus noise:
\begin{equation} 
 \begin{aligned}
d_i(\boldsymbol{\theta}) &= \sum _{\alpha} s^{\alpha}_{\nu_i}(\boldsymbol{\theta}) + n_i(\boldsymbol{\theta})\\
& = \sum _{\alpha} f^{\alpha}_i s_{\alpha}(\boldsymbol{\theta}) + n_i(\boldsymbol{\theta}) \\
    & = \sum _{\alpha} M_{i\alpha} s_{\alpha}(\boldsymbol{\theta}) + n_i(\boldsymbol{\theta}) \ ,
  \end{aligned}
\end{equation}

\noindent where $d_i$ is the sky map at frequency $\nu_i$. $M_{i \alpha}\equiv f^{\alpha}_i$ is the mixing matrix which shows the frequency dependence for the $\alpha$th component in $i$th frequency. In our analysis, $\nu_i \in \{\textrm{100, 143, 217, 353, 545, 857}\}\text{ GHz}$ and $\alpha \in \{\text{CMB }, \mathrm{tSZ}\}$. $n_i(\boldsymbol{\theta})$ is the sum of other components that are not included in $s_\alpha$ (like the CIB, whose frequency dependence is not uniform across the sky) plus instrument systematics.\par 

To extract the $\alpha$th component while nulling the other components, we need to solve the following linear equations:

 \begin{equation}
 \begin{aligned}
\sum_i c_{\alpha i} f^{\alpha}_i & = 1 \\
\sum_i c_{\alpha i} f^{\beta}_i & = 0 ,\beta \neq \alpha \ ,
\end{aligned}
\end{equation}
or more concisely:

 \begin{equation}
 \begin{aligned}
\sum_i c_{\alpha i} M _{i \beta} & = \delta _{\alpha \beta} \ .
\end{aligned}
\label{eq:ilcconstrain}
\end{equation}

\noindent $c_{\alpha i}$ is the ILC coefficient for $\alpha$ component at frequency $\nu_i$. We use $\hat{s}_{\alpha}$ to denote the estimated template for component $\alpha$. It is calculated by superposing the observed sky maps with the ILC coefficients:

 \begin{equation}
 \begin{aligned}
\hat{s}_{\alpha}(\boldsymbol{\theta}) = \sum_{i} c_{\alpha i} d_i(\boldsymbol{\theta}) = s_{\alpha}(\boldsymbol{\theta}) + \sum_i c_{\alpha i} n_i
\end{aligned}
\label{eq:estimatedi}
\end{equation}

\begin{center}
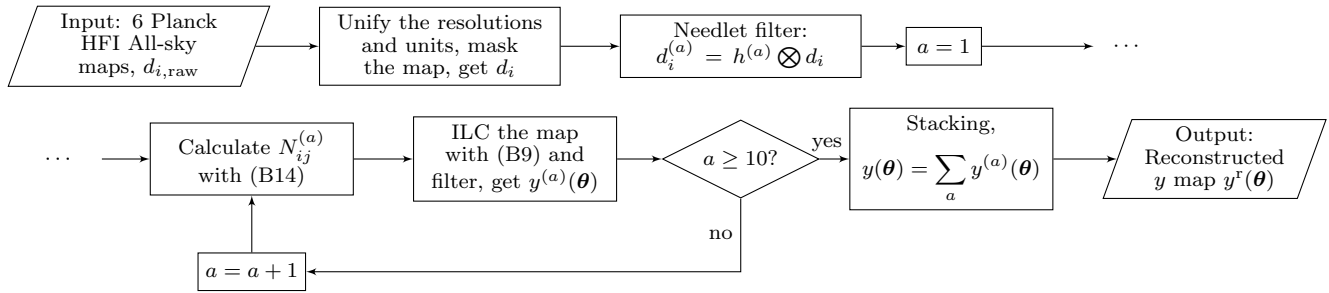
\begin{figure*}

\centering
\begin{tikzpicture}[node distance = 3cm, auto]
    \node [io,text width=2.5cm] (init) {Input: 6 Planck HFI All-sky maps, $d_{i,\text{raw}}$};
    \node [process, right of=init,text width=3cm,xshift=1cm] (preposs) {Unify the resolutions and units, mask the map, get $d_{i}$};
    \node [process, right of=preposs,text width=3cm,xshift = 1cm] (needlet) {Needlet filter: $d^{(a)}_i=h^{(a)}\bigotimes d_i$};
    \node [process, right of=needlet,xshift = -0.3cm] (inita) {$a = 1$};
    \node [continue, right of=inita,text width=0.5cm,xshift = -0.5cm] (1) {$\cdots$};
    \node [continue, below of=init,text width=0.5cm,yshift = 1.5cm,xshift = -1cm] (2) {$\cdots$};
    \node [process, right of=2,text width=2.5cm,xshift = -0.5cm] (covmat) {Calculate $N^{(a)}_{ij}$ with \eqref{eq:covmat}};
    \node [process, right of=covmat,text width=2.5cm,xshift=0.5cm] (ilcmap) {ILC the map with \eqref{eq:ilccomp} and filter, get $y^{(a)}(\boldsymbol{\theta})$ };
    \node [decision, right of=ilcmap] (decide) {$a \geq 10$?};
    \node [process, below of=covmat,yshift = 1.5cm] (app) {$a = a + 1$};
    \node [process, right of=decide,text width=2.5cm,xshift=-0.2cm] (stack) {Stacking, \[y(\boldsymbol{\theta})=\sum_a y^{(a)}(\boldsymbol{\theta})\]};
    \node [io, right of=stack,text width=2cm,xshift = 0.5cm] (output) {Output: Reconstructed $y$ map $y^{\mathrm{r}}(\boldsymbol{\theta})$};
    \path [line] (init) -- (preposs);
    \path [line] (preposs) -- (needlet);
    \path [line] (needlet) -- (inita);
    \path [line] (inita) -- (1);
    \path [line] (2) -- (covmat);
    \path [line] (covmat) -- (ilcmap);
    \path [line] (ilcmap) -- (decide);
    \path [line] (decide) -- node[anchor=south,xshift=-0.1cm] {yes} (stack);
    \path [line] (stack) -- (output);
    \path [line] (decide) |- node[anchor=east,yshift=0.5cm] {no}(app);
    \path [line] (app) -- (covmat);
\end{tikzpicture}

\begin{center}
\caption{\centering Flow chart for our {HILC} procedure.}
\end{center}

\label{fig:nilcflow}
\end{figure*}
\end{center}

In summary, we try to solve \eqref{eq:ilcconstrain} for ILC coefficients $c_{\alpha i}$, $i = 1,2,...,N_f$, where $N_f$ is the number of channels. The mixing matrix $M_{i\beta}$ is an $N_f\times N_c$ matrix, where $N_c$ denotes the number of components. For such set of linear equations, $N_c$ should be less than, or equal to $N_f$. If $N_c<N_f$, we have remaining degrees of freedom to minimize the noise residual by minimizing the $\chi ^2$:

 \begin{equation}
 \begin{aligned}
& \chi^2(\boldsymbol{\theta})  \equiv \\ & \sum_{ij} (d_i(\boldsymbol{\theta}) - \sum _{\alpha} M_{i \alpha} \hat{s}_{\alpha}(\boldsymbol{\theta})) (N^{-1})_{ij} (d_j(\boldsymbol{\theta}) - \sum _{\alpha} M_{j \alpha} \hat{s}_{\alpha}(\boldsymbol{\theta})) \ ,
 \end{aligned}
\end{equation}
\noindent where $N$ is the signal covariance matrix. Taking the partial derivative with respect to $\hat{s}_{\alpha}(\boldsymbol{\theta})$:

\begin{equation}
\frac{\partial \chi^2(\boldsymbol{\theta})}{\partial \hat{s}_{\alpha}(\boldsymbol{\theta})} = -2 \sum_{ij} M_{\alpha i} (N^{-1})_{ij} (d_j(\boldsymbol{\theta}) - \sum _{\beta} M_{j \beta}\hat{s}_{\beta}(\boldsymbol{\theta})) \ .
\end{equation}

Set it to be zero to minimize, then

\begin{equation}
\sum_{ij} M_{\alpha_i} (N^{-1})_{ij} d_j(\boldsymbol{\theta}) = \sum_{ij} M_{\alpha i} (N^{-1})_{ij} \sum _{\beta} M_{j \beta}\hat{s}_{\beta}(\boldsymbol{\theta}) \ .
\end{equation}

\noindent This leads to
\begin{equation}
\hat{s}_{\alpha}(\boldsymbol{\theta}) = \sum_{\alpha \beta k l}[(M^T N^{-1} M) ^ {-1}] _{\alpha \beta} M_{\beta k} (N^{-1})_{kl} d_l(\boldsymbol{\theta}) \ .
\label{eq:ilccomp}
\end{equation}

\noindent Comparing with Eq \eqref{eq:estimatedi}, the coefficient for component $\alpha$ at frequency channel $i$ is

\begin{equation}
c_{\alpha i} = \sum_{\alpha \beta k}[(M^T N^{-1} M) ^ {-1}] _{\alpha \beta} M_{\beta k} (N^{-1})_{ki} \ .
\label{eq:ilccoef}
\end{equation}

It is straightforward to confirm that \eqref{eq:ilccoef} satisfiles \eqref{eq:ilcconstrain}. Another method is to use Lagrange multiplier to find the minimum of $\chi$ subject to constraints \eqref{eq:ilcconstrain}. Both methods give the same result \eqref{eq:ilccoef}.

The frequency dependence for each component is contained in the mixing matrix $M_{i \alpha}$. Free-free scattering and synchrotron have a decreasing frequency spectrum in HFI region and are significant only at the low frequency, so we ignore them here. For our fiducial reconstructed $y$ map we only project out the CMB components. The other components and noises are minimized. The units of raw sky maps are thermal temperature $\mathrm{K_{CMB}}$, so the dependence for each component is as following:\par

Intensity of primary CMB fluctuation is a black body spectrum with monopole temperature 2.725K \citep{fixsen2009temperature}. In $\mathrm{K_{CMB}}$ unit, the CMB signal is independent of frequency so $f^{\mathrm{CMB}}(\nu)=1$ for all channels, thus we have:

\begin{equation}
s_{\nu}^{\text{CMB}}(\boldsymbol{\theta}) = s_{\text{CMB}}(\boldsymbol{\theta}) \ ,
\end{equation}
where $s^{\text{CMB}}(\boldsymbol{\theta})$ is the CMB template which depends only on position $\boldsymbol{\theta}$. \par

For the tSZ signal \citep{birkinshaw1999sunyaev}:
\begin{equation}
s_{\nu}^{\text{tSZ}}(\boldsymbol{\theta}) = S_{\text{tSZ}}(x) T_{\mathrm{CMB}} y(\boldsymbol{\theta}) \ ,
\end{equation}
where $x \equiv h\nu/k_{\mathrm{B}}T_{\text{CMB}}$. $S_{\mathrm{tSZ}}$ is defined in \eqref{eq:def_yparam} and $T_{\mathrm{CMB}}=2.725\mathrm{K}$ is the monopole temperature of the CMB. The Compton-y parameter $y(\boldsymbol{\theta})$ is a dimensionless parameter describing the spatial dependence of the tSZ effect. So the $y$ map reconstruction is to find ILC coefficients $c_{\mathrm{tSZ},i}$ using \eqref{eq:ilccoef}.

\subsection{Reconstruct tSZ Map}
 \label{subs:ilc}
 
In practice, before taking ILC procedure, we need to prepocess the raw temperature maps as follows:\\

 1. Convert all maps to $\mu \mathrm{K_{CMB}}$ with the conversion coefficients provided in the Planck explanatory supplement\footnote{We used the "545-avg" and "857-avg" values from \url{https://wiki.cosmos.esa.int/planckpla2015/index.php/UC_CC_Tables}}.\par

 2. The corresponding angular resolutions are FWHM$_0$ = \{9.66, 7.27, 5.01, 4.86, 4.84, 4.63\} arcmin. To first order we can take the Planck beam function to be Gaussian \citep{planck2014planckbeam}. We smooth the maps to a common angular resolution of 10 arcmin by convolving each map with a Gaussian beam with $\text{FWHM} = \sqrt{10^2 - \text{FWHM}_0^2}$ arcmin.\par

3. Cut the sky with $y$ mask described in Section \ref{section:map_method}

We use the {Harmonic} Internal Linear Combination (HILC) method to generate the $y$ map. {The ILC procedure is taken in harmonic space.} The raw band map is first transformed into $\ell$ space and multiplied by a needlet filter $\{h^{(a)}(\ell) \}$ then transformed back into real space. The output map is called a needlet-filtered map. $h^{(a)}(\ell)$ centeres at a certain scale $\ell_a$, so a needlet-filtered map corresponding to $h^{(a)}(\ell)$ preserves intensity around scale $\ell_a$. We make ILC maps independently for each needlet window, and noise is minimized independently for different angular scales.\par 

\begin{figure*} \centering

\label{fig:ymap_this}
\includegraphics[width=0.94\textwidth]{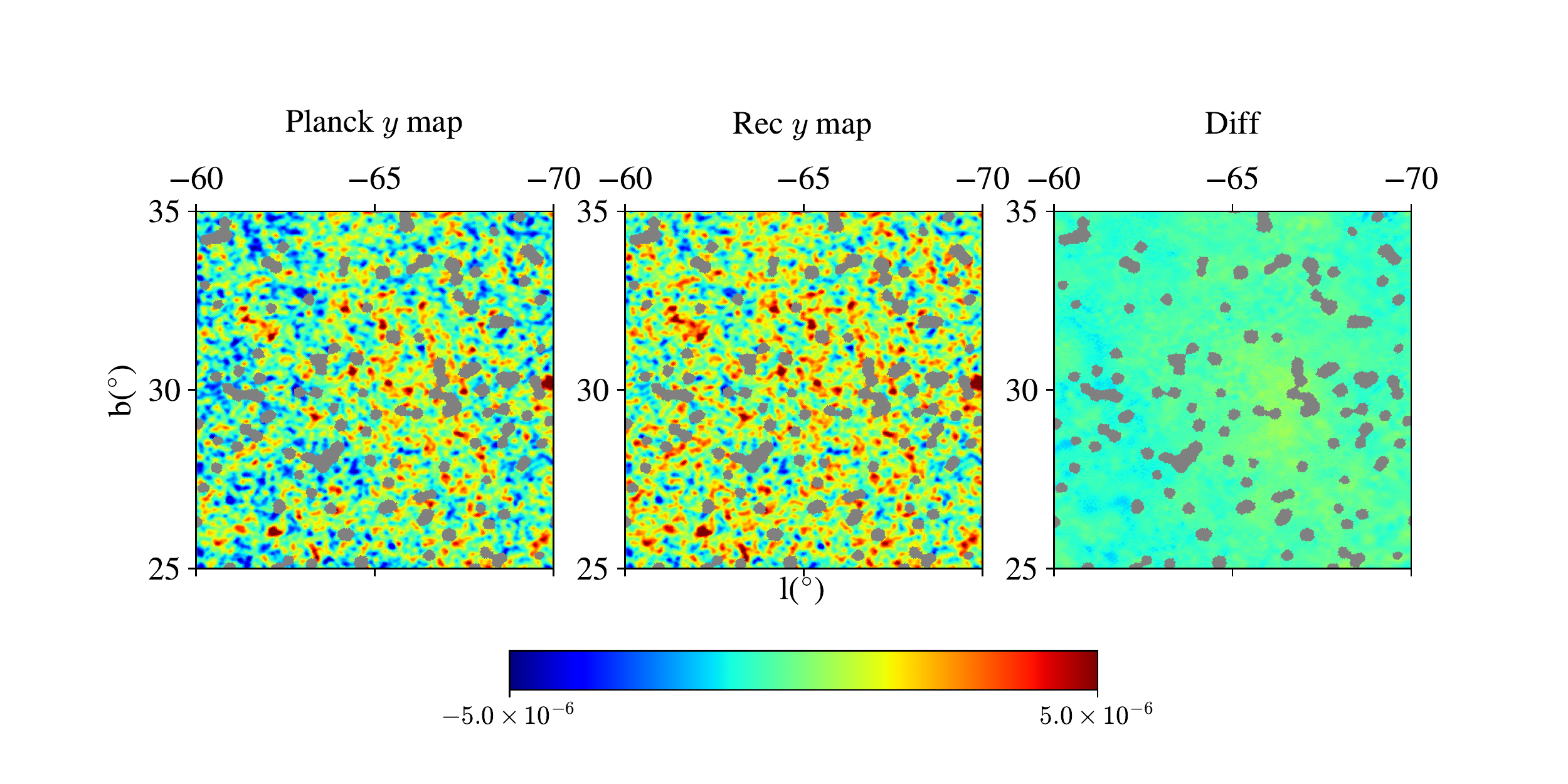}

\begin{center}
\caption{ $y$ signal of a small region of the sky for Planck NILC map and our $y$ map. The right-most panel shows the difference.}
\end{center}
\label{fig:ymap_compare}
\end{figure*}

Based on \citet{basak2012needlet}, we use 10 Gaussian window functions peaking at different scales as $\{h^{(a)}(\ell) \}$. These needlet windows are also used in \citet{planck2014plancksz} and \citet{aghanim2016planckcib} in their NILC procedures. Appendix \ref{app:nilc} contains the details about the needlet windows. As preprocessing, we filter the 6 raw band maps with these 10 needlet filters to make 10 sets of band maps $d^{(a)}_i,1 \leq a \leq 10$ each corresponding to a needlet window. Each set has 6 band maps at 6 HFI channels. We mask the maps after needlet filtering in order to prevent ringing effects. \par 

The ILC is performed independently with each set of filtered maps. We first calculate the covariance matrix of each Needlet-filtered sky map:\par 

\begin{equation}
N_{ij}^{(a)} = \left\langle d_i^{(a)}(\boldsymbol{\theta}) d_j^{(a)}(\boldsymbol{\theta}) \right\rangle_{\boldsymbol{\theta} \in D} \ ,
\end{equation}

\noindent where $a$ is the needlet index. $D$ is the domain the real space we are interested in, typically a masked map. In practice, the covariance matrix is estimated by multiplying together signals of the same pixel in $i$th and $j$th map, then summing over pixels in the domain $D$.

\begin{figure}\centering

\includegraphics[width=0.9\linewidth]{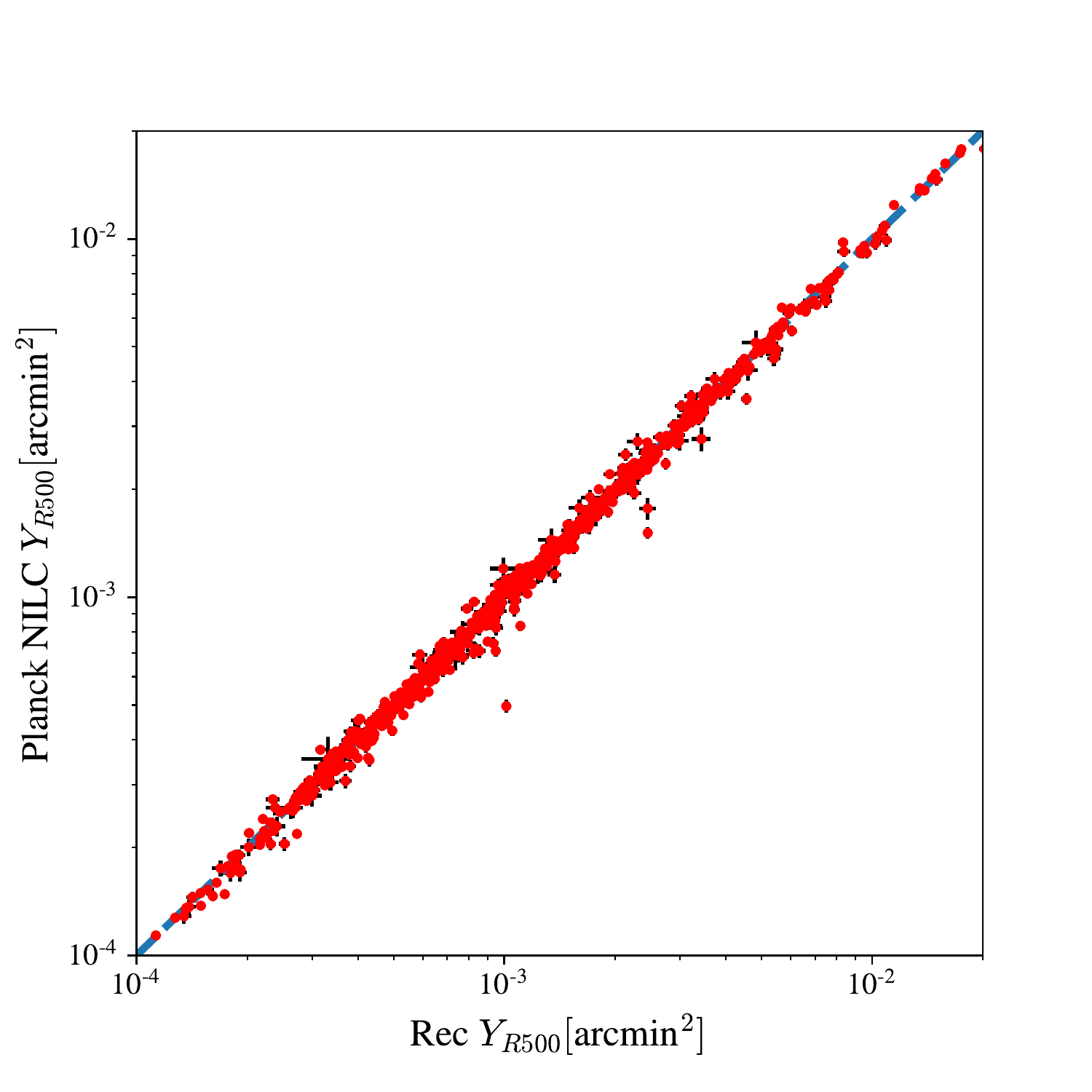}

\caption{ Comparison between the measured tSZ flux of the Planck cluster sample measured in Planck NILC map and our reconstructed $y$ map.}
\label{fig:ycluster}
\end{figure}

\begin{equation}
\hat{N}_{ij}^{(a)} = \frac{1}{N_p}\sum_{p \in D} d_i^{(a)}(p) d_j^{(a)}(p) \ ,
\label{eq:covmat}
\end{equation}
where $N_p$ is the number of pixels in domain $D$. For our analysis, $D$ is the unmasked sky which is discussed above. 

The weight for component separation is calculated independently for each needlet window. Thus we can make component maps for each scale. The reconstructed tSZ map is obtained by re-applying the needlet windows $h^{(a)}(\ell)$ and co-adding tSZ maps in all needlet windows. Our {HILC} pipeline is summarized as a flow chart in Fig.\ref{fig:nilcflow}. {Our HILC method differs from NILC method in that the domain $D$ in \eqref{eq:covmat} includes all unmasked pixels in the map, while for NILC $D$ depends on scale.} Another difference between our reconstructed $y$ map (labeled as $y^{\textrm{r}}$ hereafter) and Planck NILC $y$ map (labeled as $y^{\textrm{p}}$ hereafter) is that we only use 6 HFI maps while Planck NILC uses LFI maps at large angular scales. The Planck NILC map only masked the most central part of the Milky Way, which is about 2\% of the sky. This will likely bring up residual galaxy signals but they should not affect the cross-correlations evidently.\par  

{In summary, the $y$ map reconstruction is formulated as:}

\begin{equation}
    {y^{\mathrm{r}}(\boldsymbol{\theta}) =
    \sum_{a}y^{(a)}(\boldsymbol{\theta}) 
     = \sum_{i,a}c^{(a)}_{i}d^{(a)}_i(\boldsymbol{\theta})}
\end{equation}{}
{where $y^{(a)}$ is the $y$ map in the $a$th needlet window, $c^{(a)}_i$ is the ILC coefficient for the $a$th needlet window and the $i$th frequency channel.}

Fig.\ref{fig:ymap_compare} shows the  $y$  signal in the same field for $y^{\textrm{r}}$ and $y^{\textrm{p}}$ and their difference. Both maps agree with each other well except for some large-scale difference due to galactic residual. We also calculate integrated  $y$  signal within $R_{500}$ for 858 Planck tSZ clusters \citep{planck2014plancksz} on Planck $y$ map and our reconstructed $y$ map. The signal-to-signal scatter plot is shown in Fig.\ref{fig:ycluster}. From Fig.\ref{fig:ycluster} we can see that the  $y$  signal from both maps agree well with each other. A paired Student t-test shows that the tSZ flux in our map agree with that from Planck NILC map to a confidence level of 7$\sigma$. The difference is due to the different ILC model and covariance matrices.

{We take the calibration uncertainty into account. We calculate the uncertainty that propagates into $y^{\mathrm{r}}$ by sampling 20 sets of sky maps with random calibration factors:}

\begin{equation}
{
    y^{\mathrm{r}}_{\mathrm{sample}} = \sum_{i,a}(c^{(a)}_{i}\epsilon_id^{(a)}_i)}
    \label{eq:deltay_calib}
\end{equation}
{where $\epsilon_i$ a Gaussian random number centered at 1 with standard deviation equals to calibration uncertainties from Table 6 of \citet{adam2016planck_inst}. The uncertainty in $y^{\mathrm{r}}$ is the standard deviation of these 20 $y^{\mathrm{r}}_{\mathrm{sample}}$.} 

\subsection{CIB subtracted y map}
\label{subsec:cib-sub}

The Planck collaboration made 3 CIB maps in 353GHz, 545GHz, and 857GHz \citep[][]{aghanim2016planckcib} by disentangling the CIB signal from a galactic dust emission map. The galactic dust emission map is generated with a Generalized ILC method using all the 9 Planck all-sky maps. The CIB covariance matrix is acquired from simulated CIB maps \citep[][]{ade2011planckcib}. The units of the maps are MJy/sr and their angular resolution is 5 arcmin. \par

Our reconstructed $y$ map can be decomposed as a combination of true $y$ signal and a superposition of error terms:

\begin{equation}
\begin{aligned}
y^{\mathrm{r}}(\boldsymbol{\theta}) & = \sum_{i,a}c^{(a)}_{i}d^{(a)}_i(\boldsymbol{\theta})\\
& =y^{\mathrm{true}}(\boldsymbol{\theta}) + \sum_{i,a}c^{(a)}_{i}n^{(a)}_i(\boldsymbol{\theta})\\
& = y^{\mathrm{true}}(\boldsymbol{\theta}) + \sum_{i,a}c^{(a)}_{i}s^{\text{CIB},(a)}_i(\boldsymbol{\theta}) + \sum_{i,a}c^{(a)}_{i}n'^{(a)}_i(\boldsymbol{\theta}) \ ,
\end{aligned}
\label{eq:estmymap}
\end{equation}

\noindent where, in the last line, we single out the residual CIB contributions to the $y$ map explicitly. Here $(a)$ is the needlet index. $c^{(a)}_{\nu}$ is the ILC coefficient for  $y$  (we omit the component index $\alpha$ in \eqref{eq:ilccoef} because we are only concerned about  $y$  now). $y^{\mathrm{r}}$ is the reconstructed  $y$  signal and  $y^{\mathrm{true}}$  is the true  $y$  signal. The CMB is removed based on its known spectrum while noise is minimized but not completely removed. The noise term $n_i$ contains both CIB signal $s^{\text{CIB}}_i$ and other noise $n'_i$ both from the sky (CO emission) and from the instrument (photon noise). cross-correlating both side of \eqref{eq:estmymap} with $\kappa$, we get:

\begin{equation}
\begin{aligned}
C^{\kappa\times  y^{\mathrm{r}}}_{\ell} = C^{\kappa\times   y^{\mathrm{true}}}_{\ell} &+ \sum_{i,a} c^{(a)}_{i} h^{(a)}(\ell) C^{\kappa\times  s^{\text{CIB}}_{\nu_i}}_{\ell} \\
& + \sum_{i,a} c^{(a)}_{i} h^{(a)}(\ell) C^{\kappa\times  n'_i}_{\ell} \ ,
\end{aligned}
\end{equation}

\noindent where $C^{\kappa\times  y^{\mathrm{r}}}_{\ell}$ can be directly measured from the $y^{\mathrm{r}}$ and $\kappa$ maps. It consists of the true $\kappa\times   y^{\mathrm{true}}$ signal as well as contamination from CIB and other noise.

To correct the CIB contamination, we make a CIB\mbox{-}subtracted  $y$ map (denoted as $y^{\mathrm{c}}$ hereafter). We first make CIB\mbox{-}subtracted temperature maps and then perform the same {HILC} procedure:

\begin{equation}
\begin{aligned}
y^{\mathrm{c}}(\boldsymbol{\theta}) & = \sum_{i,a}c^{(a)}_{i}(d^{(a)}_i(\boldsymbol{\theta}) - s^{\text{CIB},(a)}_{i}(\boldsymbol{\theta}))\\
& = y^{\mathrm{true}}(\boldsymbol{\theta}) + \sum_{i,a}c^{(a)}_{i}n'^{(a)}_i(\boldsymbol{\theta}) \ .
\end{aligned}
\label{eq:ycibsubtracted}
\end{equation}

We take the Planck CIB maps as $s^{\text{CIB},(a)}_{i}$ at 353, 545, 857 GHz. Although the CIB signal is low in 100,143, and 217GHz, the corresponding ILC coefficients is higher than for high frequencies, so the CIB cannot be ignored for these frequencies. A correlation coefficient calculation among the three CIB maps shows that they are correlated to a level of 0.994, which means they are nearly proportional to each other. Therefore we make 3 model CIB maps at 100,143, and 217GHz by scaling Planck 353GHz CIB map with a homogeneous CIB spectrum model in order to efficiently subtract CIB signal from the raw temperature maps.  Based on \citet{schmidt2014inferring}, the redshift distribution of CIB sources peaks at around $z\approx1.2$ which is independent of frequency. According to the CIB model parameters given in Table.\ref{table:cibmodel}, the connecting frequency between grey-body and powerlaw region $\nu_0$ is about 4400GHz at $z=1.2$. So $\Theta[\nu,T_d]$ in the Planck HFI bands is a greybody spectrum $\Theta[\nu_i,T_d(z)]$.  Therefore we use $\Theta[\nu,T_d(1.2)]$ to model the frequency dependence of the CIB in 100,143, and 217GHz:

\begin{equation}
s^{\mathrm{CIB}}_i = s^{\mathrm{CIB}}_{\mathrm{353GHz}}\frac{\Theta[\nu_i,T_c(1.2)]}{\Theta[\mathrm{353GHz},T_c(1.2)]} \ ,
\end{equation}
where $\Theta[\nu,T_c]$ is defined in Eq \eqref{eq:cibspectrum}. Values of $\beta$, $T_{c0}$ are from Table \ref{table:cibmodel}. We smooth all the six CIB maps to an angular resolution of 10 arcmin and subtract them from the raw sky maps to make CIB\mbox{-}subtracted sky maps. We then carry out the same {HILC} procedure with these sky maps to make $\hat{y}^{\textrm{CIB\mbox{-}subtracted}}$.\par

\subsection{Dust Nulled $y$ Map}
\label{subsec:dust-nulled}
\begin{figure} \centering
\includegraphics[width=\columnwidth]{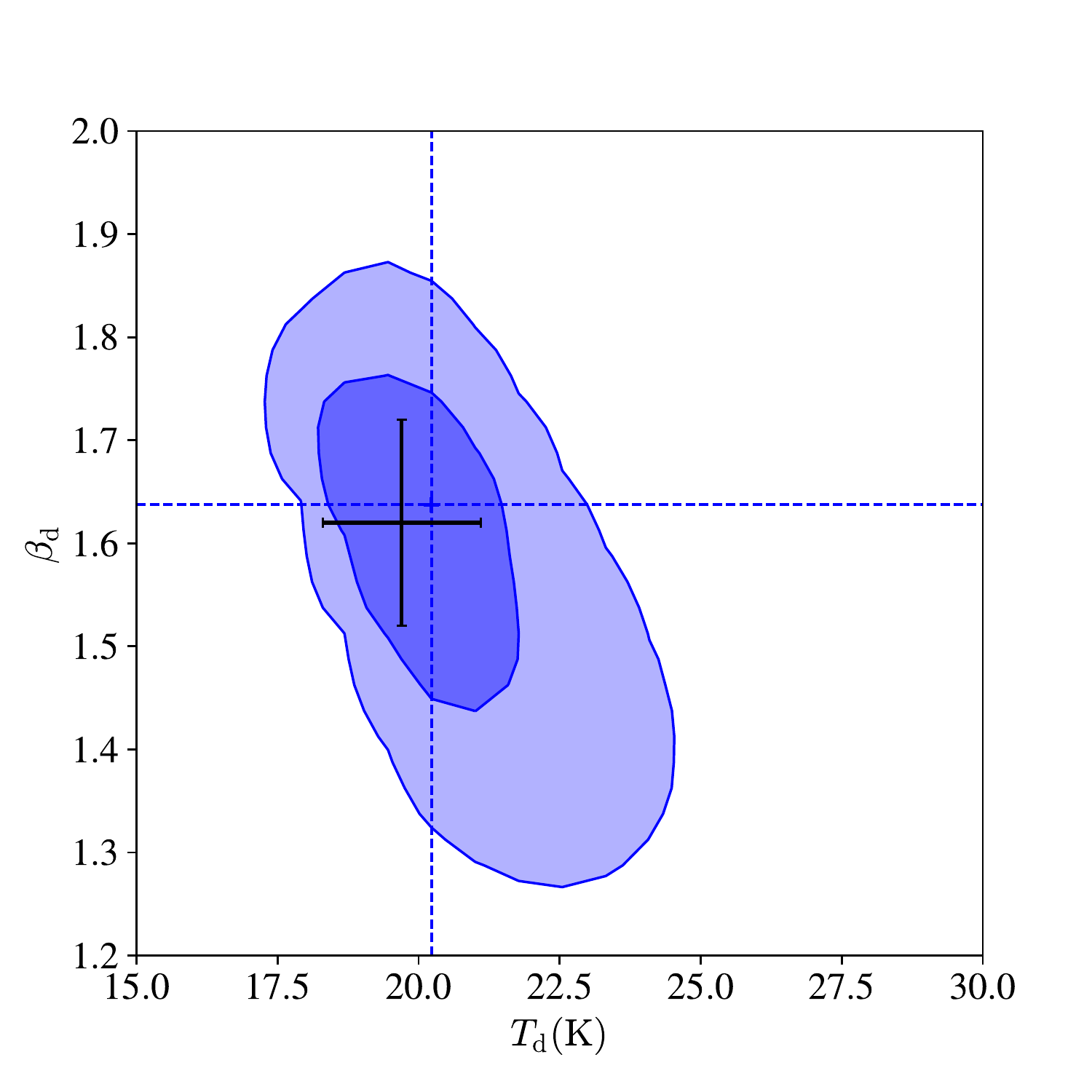}

\caption{2-D Histogram for $T_d$ and $\beta_d$ in our $y$ footprint. The dust model we use here is the Planck COMMANDER thermal dust map \citep{adam2016planckdust}. Contours are the 1,2$\sigma$ levels. Dashed lines show the mode value of $T_d$ and $\beta_d$. Black cross is the estimated mean values and standard deviation by Planck Collaboration. }
\label{fig:dustmodel}
\end{figure}

The intensity of thermal galactic dust can be modeled as a grey body spectrum \citep{ade2016planck}:
\begin{equation}
I_{\nu}^{\text{dust}}(\boldsymbol{\theta}) \propto \nu ^ {\beta_\text{d}} B_{\nu}(T_\text{d}) I_{\text{dust}}(\boldsymbol{\theta}) \ ,
\end{equation}
where $T_\text{d}$ is the dust temperature and $\beta_\text{d}$ is the dust spectral index. So the dust signal in $\mathrm{K_{CMB}}$ is :

\begin{equation}
s_{\nu}^{\text{dust}}(\boldsymbol{\theta}) \propto \nu ^ {\beta_d-2} B_{\nu}(T_d) s_{\text{dust}}(\boldsymbol{\theta}) \ .
\end{equation}

$\beta_\text{d}$ and $T_\text{d}$ both vary across the sky, so it is not valid to write $s^{\mathrm{dust}}_{\nu}(\boldsymbol{\theta})=f^{\mathrm{dust}}_{\nu}s_{\mathrm{dust}}(\boldsymbol{\theta})$. We can formally decompose dust according to different combination of $\beta_\text{d}$ and $T_\text{d}$, each corresponding to a 'dust component'. There are infinite and uncountable combinations, so it is impossible to use ILC to project the dust out completely with a finite number of frequency channels. In principle, we can project out at most 4 dust components with 6 band maps (the other 2 are used to null the CMB and preserve $y$), but then we run out of degree of freedom to minimize the variance. The residue of other dust components and noises will thus dominate the output $y$ map. 

In the Planck $y$ maps, our reconstructed $y$ map and our CIB\mbox{-}subtracted $y$ map, the dust signal is suppressed but not projected out. Since the dust signal is originated in the Milky Way, it should not correlate with either $\kappa$ or $\phi$. But it can affect the error in $\kappa\times y$ and $\phi\times y$ signal and thus affect the robustness of cross-correlation conclusions. 

To test it, we make a set of $y$ maps with different dust residuals. We include one dust component (i.e. one combination of \{$\beta_\text{d}$, $T_\text{d}$\}) in the mixing matrix $M_{i\alpha}$. By varying $\beta_d$ and $T_d$ we change the dust component to be projected out. Different set of \{$\beta_\text{d}$, $T_\text{d}$\} gives $y$ map with different dust residuals. In this work we choose five $\beta_\text{d}$ values from 1.3 to 1.9 and make five $y$ maps. As in subsection \ref{subsec:cib-sub}, CIB signal is subtracted out before producing the $y$ map.

The spatial distribution of \{$\beta_\text{d}$, $T_\text{d}$\} is given in a dust model map made by the Planck collaboration. The values are measured to be $T_d=19.7\pm0.4$K, $\beta_d=1.62\pm0.1$. The 2-d histogram of \{$\beta_\text{d}$, $T_\text{d}$\} (Fig.\ref{fig:dustmodel}) in the footprint of $\hat{y}^{\mathrm{rec}}$ map shows mode values of $\beta_\text{d} = 1.64$, $T_\text{d} = 20.23\text{K}$. Dust components with these parameter values contribute most in our $y$ footprint.

\section{The Needlet Windows}
\label{app:nilc}
\begin{figure}\centering

\includegraphics[width=\linewidth]{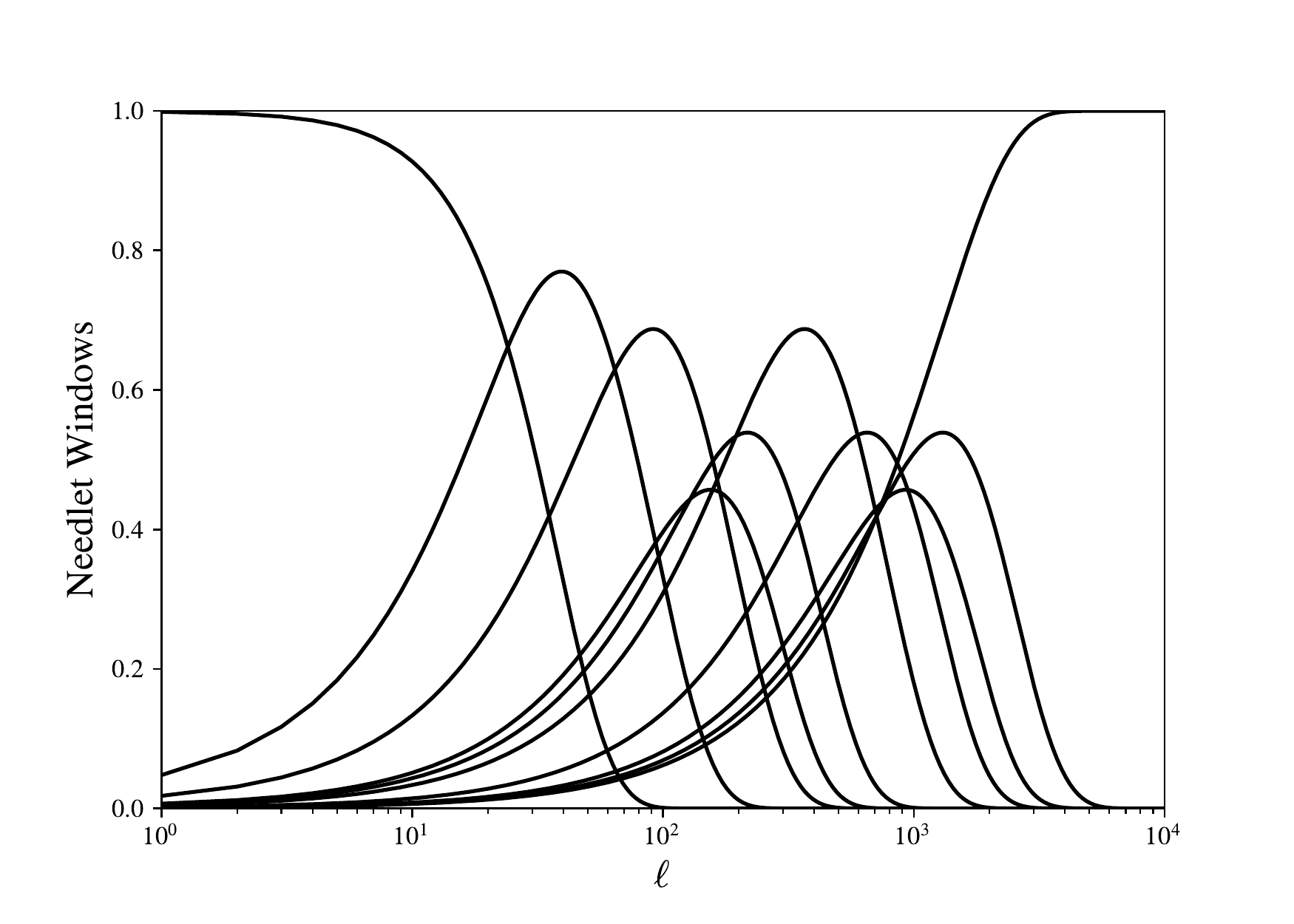}

\caption{Needlet windows $h^{(a)}(\ell)$ acting as bandpass filters in $\ell$ space.}
\label{fig:needlet}
\end{figure}

Our ILC process is performed in a Needlet frame. Needlet is first introduced by \citet{narcowich2006localized} as a particular construction of a wavelet	frame on a sphere. The most  distinctive  property  of  the  needlets  is  their  simultaneous perfect localization in the spherical harmonic domain (actually they are spherical polynomials) and potentially excellent localization in the spatial domain.

Basically, the raw temperature maps are first filtered into needlet windows by first make spherical harmonic transforms of the maps $x_{\ell m}$, then multiplied by the needlet window $h^{(a)}(\ell)$ and transformed back into real space. The result is called a needlet map, characterized by a given range of angular scales given in $h^{(a)}(\ell)$. ILC is performed for each needlet scale, and the synthesized map is obtained by co-adding the ILC estimates for each needlet scale. In this work,the needlet bandpass windows are defined following \citep{aghanim2016planckcib}, which is a set set of successive Gaussian beam transfer functions in harmonic space.

\begin{equation}
\begin{aligned}
h^{(1)}(\ell) &= \sqrt{b_1(\ell)^2}, \\
h^{(a)}(\ell) &= \sqrt{b_{a}(\ell)^2 - b_{a-1}(\ell)^2}, 1<a<10 \\
h^{(10)}(\ell) &= \sqrt{1 - b_{9}(\ell)^2} \ ,
\end{aligned}
\end{equation}
where

\begin{equation}
b_a(\ell) = \exp \left(-\ell(\ell+1)\sigma^2_a/2\right) \ ,
\end{equation}
and

\begin{equation}
\sigma_a = \left(\frac{1}{\sqrt{8\ln2}}\right)\left(\frac{\pi}{180\times60'}\right)\text{FWHM}[a]
\end{equation}
with FWHM = $[300',120',60',45',30',15',10',7.5',5']$. So we have

\begin{equation}
\sum^{10}_{a=1}\left(h^{(a)}(\ell)\right)^2 = 1 \ .
\end{equation}

So the signal for the output synthesized map from different needlet is conserved.

To calculate the needlet-filtered map $d^{(a)}_i$, we first calculate the spherical harmonic transformation of $d_i$:

\begin{equation}
d_i(\boldsymbol{\theta}) = \sum_{\ell m} x_{i,\ell m} Y_{\ell m}(\boldsymbol{\theta}) \ .
\end{equation}

$x_{i,\ell m}$ is the spherical harmonic coefficient for map of the $i$th channel. Multiply it by the needlet filter $h^{(a)}(\ell)$ and transform back, we get the needlet-filtered map:

\begin{equation}
d_i^{(a)}(\boldsymbol{\theta}) = \sum_{\ell m} h^{(a)}(\ell) x_{i,\ell m} Y_{\ell m}(\boldsymbol{\theta}) \ .
\label{eq:defdai}
\end{equation}
\vspace{0.2cm}

\section{tSZ residual in the CMB Lensing Map}
\label{app:sys_cmblens}

{\citet{PhysRevD.98.023534}} and \citet{chen2018impact} pointed out that the tSZ residual in the CMB map is not negligible when doing CMB-large scale structure cross-correlation. The Planck 2015 CMB maps contain tSZ residual by construction. Planck 2018 data release includes a tSZ-cleand CMB map constructed by SMICA method \citep{aghanim2018planck_param}, and the lensing map used in this paper is made from it.

Since the $\phi$ map is made out of the CMB map, the tSZ residual could propagate to $\phi$ map and contaminate the $\phi\times y$ signal. We test this contamination by cross-correlation $y$ map with different $\phi$ maps: with and without deprojecting the tSZ signal.

\begin{figure}
\centering

\includegraphics[width=\columnwidth]{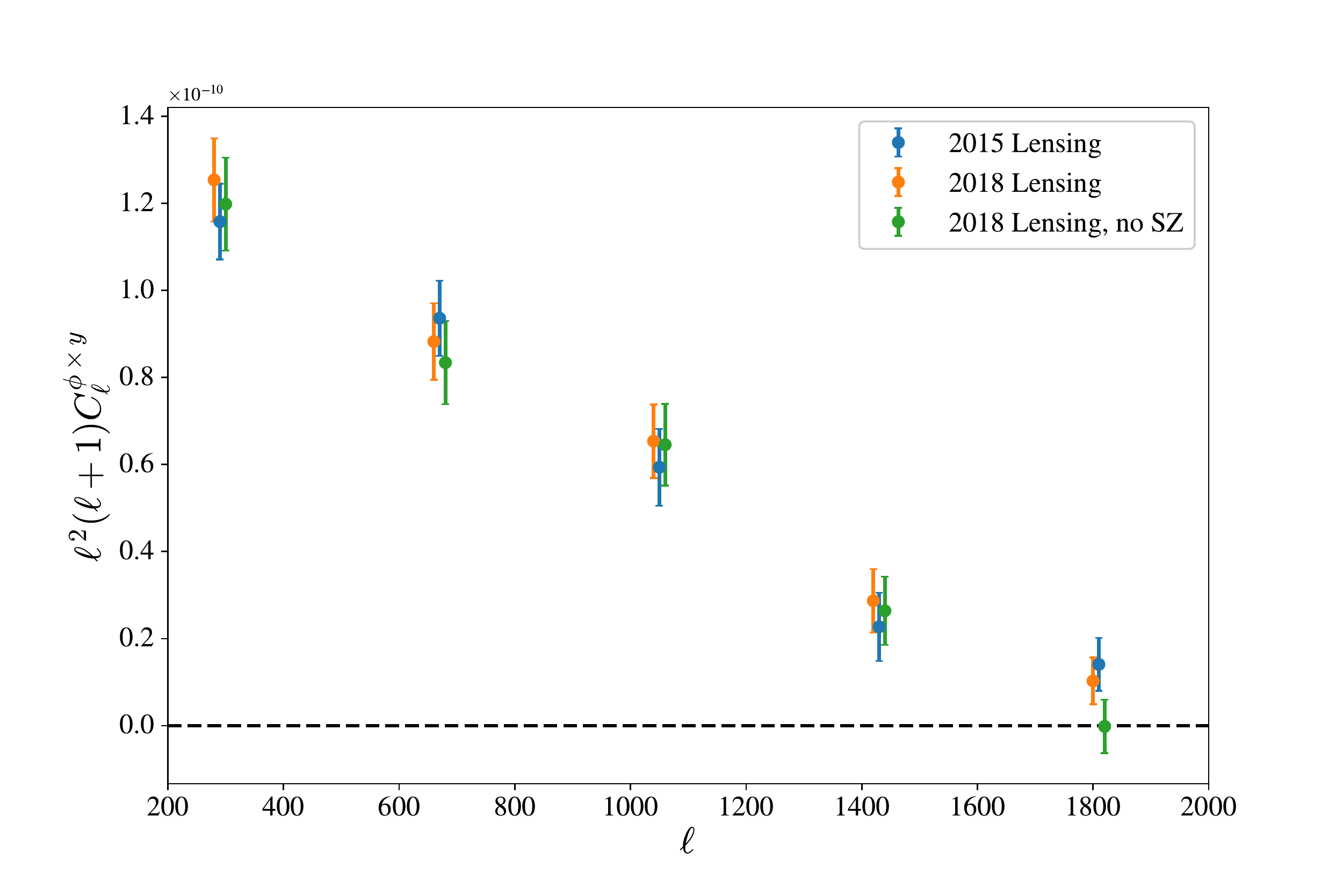}

\caption{$y\times \phi$ with different CMB lensing data. The $y$ map used here is the Planck NILC $y$ map.}
\label{fig:py_nosz}
\end{figure}

\ref{fig:py_nosz} shows that the tSZ residual in the CMB lensing map does not affect $y\times \phi$ signal significantly, although, as is discussed in \citet{chen2018impact}, tSZ residual in the CMB map does contribute to CMB-LSS measurements. A similar analysis like Fig.\ref{fig:phiy} with the other two lensing maps shows consistent CIB contamination estimations.

\end{document}